\documentclass[acmsmall,screen]{acmart}

\setcopyright{none}
\acmYear{2025}
\acmDOI{XXXXXXX.XXXXXXX}

\acmJournal{JACM}
\acmVolume{1}
\acmNumber{1}
\acmArticle{1}
\acmMonth{10}

\usepackage{booktabs}
\usepackage{tabularx}
\usepackage{amssymb}
\usepackage{pifont}
\usepackage{makecell}
\usepackage{subfigure}
\usepackage{multicol}
\usepackage{multirow}
\usepackage{overpic}
\usepackage{tcolorbox}
\usepackage{enumitem}
\usepackage{wrapfig}

\newcommand{\cmark}{\ding{51}} % 定义对勾符号
\newcommand{\xmark}{\ding{55}} % 定义叉号符号

\begin{document}

\title{Deep Learning-based Intrusion Detection Systems: A Survey}

\author{Zhiwei Xu}
\orcid{0000-0003-4430-3727}
\author{Yujuan Wu}
\orcid{0009-0009-9292-4860}
\author{Shiheng Wang}
\orcid{0009-0008-9088-7827}
\author{Jiabao Gao}
\orcid{0009-0008-6544-9382}
\author{Tian Qiu}
\orcid{0000-0003-2005-1326}
\author{Ziqi Wang}
\orcid{0000-0002-0976-5128}
\author{Hai Wan}
\orcid{0000-0002-9608-5808}
\author{Xibin Zhao}
\orcid{0000-0002-6168-7016}
\authornote{Xibin Zhao is the corresponding author.}
\affiliation{
  \institution{KLISS, BNRist, School of Software, Tsinghua University}
  \city{Beijing}
  \country{China}
}
\email{zxb@tsinghua.edu.cn}

\begin{abstract}
Intrusion Detection Systems (IDS) have long been a hot topic in the cybersecurity community. In recent years, with the introduction of deep learning (DL) techniques, IDS have made great progress due to their increasing generalizability. The rationale behind this is that by learning the underlying patterns of known system behaviors, IDS detection can be generalized to intrusions that exploit zero-day vulnerabilities. In this survey, we refer to this type of IDS as DL-based IDS (DL-IDS). From the perspective of DL, this survey systematically reviews all the stages of DL-IDS, including data collection, log storage, log parsing, graph summarization, attack detection, and attack investigation. To accommodate current researchers, a section describing the publicly available benchmark datasets is included. This survey further discusses current challenges and potential future research directions, aiming to help researchers understand the basic ideas and visions of DL-IDS research, as well as to motivate their research interests.
\end{abstract}

%% The code below is generated by the tool at http://dl.acm.org/ccs.cfm. Please copy and paste the code instead of the example below.



% \received{20 February 2007}
% \received[revised]{12 March 2009}
% \received[accepted]{5 June 2009}

\maketitle

\section{Introduction}

The promising Internet of Everything connects people, processes, data, and things through the Internet \cite{evans2012internet}, bringing convenience and efficiency to the world. Yet its inevitable security vulnerabilities could be exploited by deliberate attackers. With increasingly sophisticated attack methods such as Advanced Persistent Threat (APT), the attackers are in a threatening position to sabotage network systems or steal sensitive data. The detection of intrusions, particularly based on DL, has consequently been a prominent topic in the cybersecurity community. 

The automated system for detecting intrusions is known as IDS. The limitations of IDS may result in terrible damage to enterprises. One example is the recent Colonial Pipeline Ransomware Attack \cite{beerman2023review}. In April 2021, the hacking group DarkSide launched a ransomware attack on Colonial Pipeline, the biggest oil pipeline company in the United States, using an unused VPN account. Due to this attack, 5,500 miles of transportation pipelines were forced to shut down, affecting nearly 45\% of the fuel supply on the Eastern Coast. The Colonial Pipeline paid \$4.4 million ransom money, in addition to the theft of over 100 GB of data. If the malware intrusion can be detected in time, the influence of this attack can be greatly mitigated or even eliminated.

\subsection{Tough but Bright Intrusion Detection System}

\begin{figure}[t]
    \centering
    \subfigure[Trend of CVE records and IDS papers.]{
    \begin{minipage}{0.505\linewidth}
        \centering
        \includegraphics[width=\linewidth, trim=0 45 0 0]
        {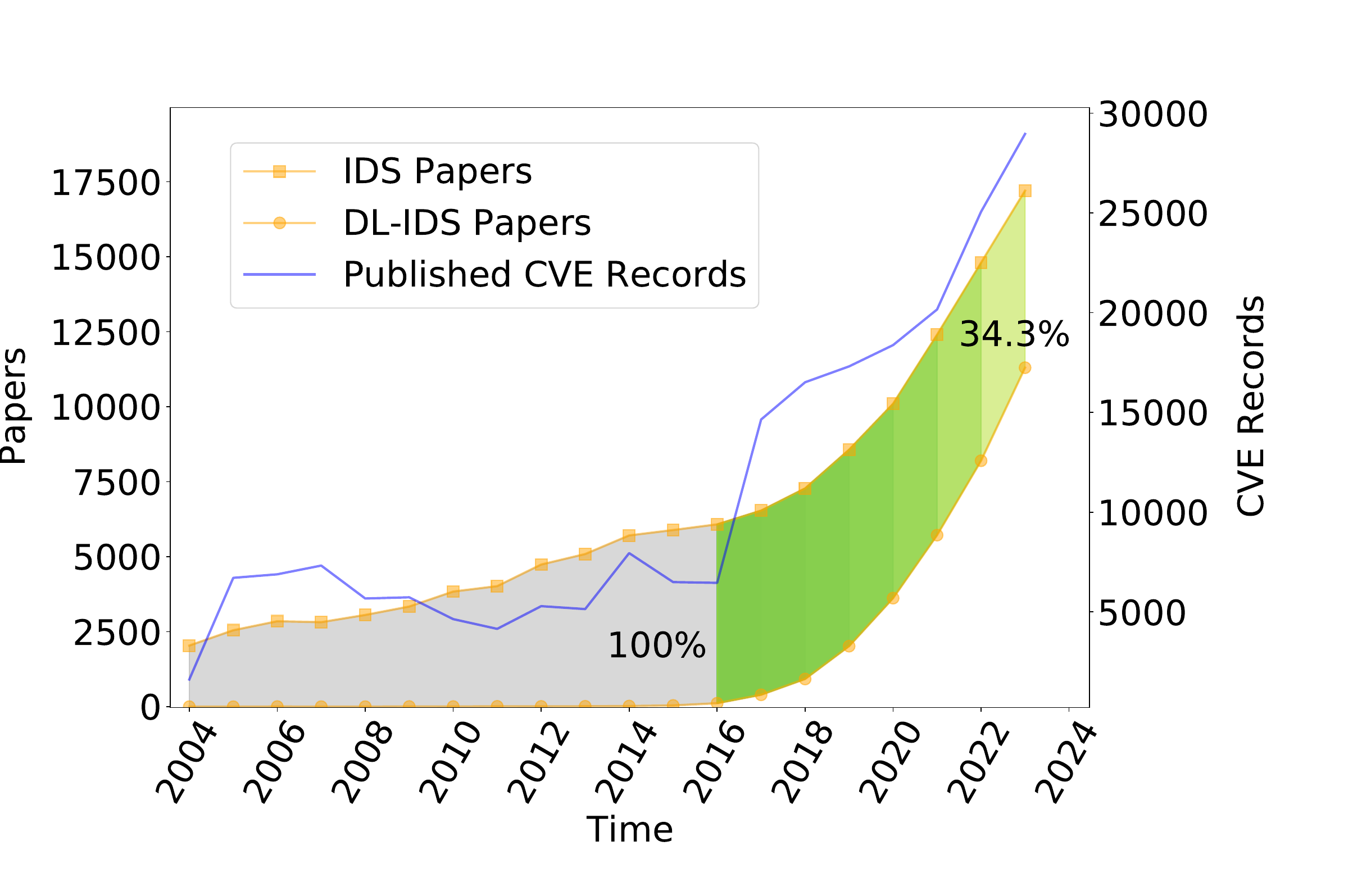}
        \label{fig:cve}
    \end{minipage}
    }
    \subfigure[Category of CNNVD vulnerabilities.]{
    \begin{minipage}{0.46\linewidth}
        \centering
        \includegraphics[width=\linewidth, trim=0 0 0 0]{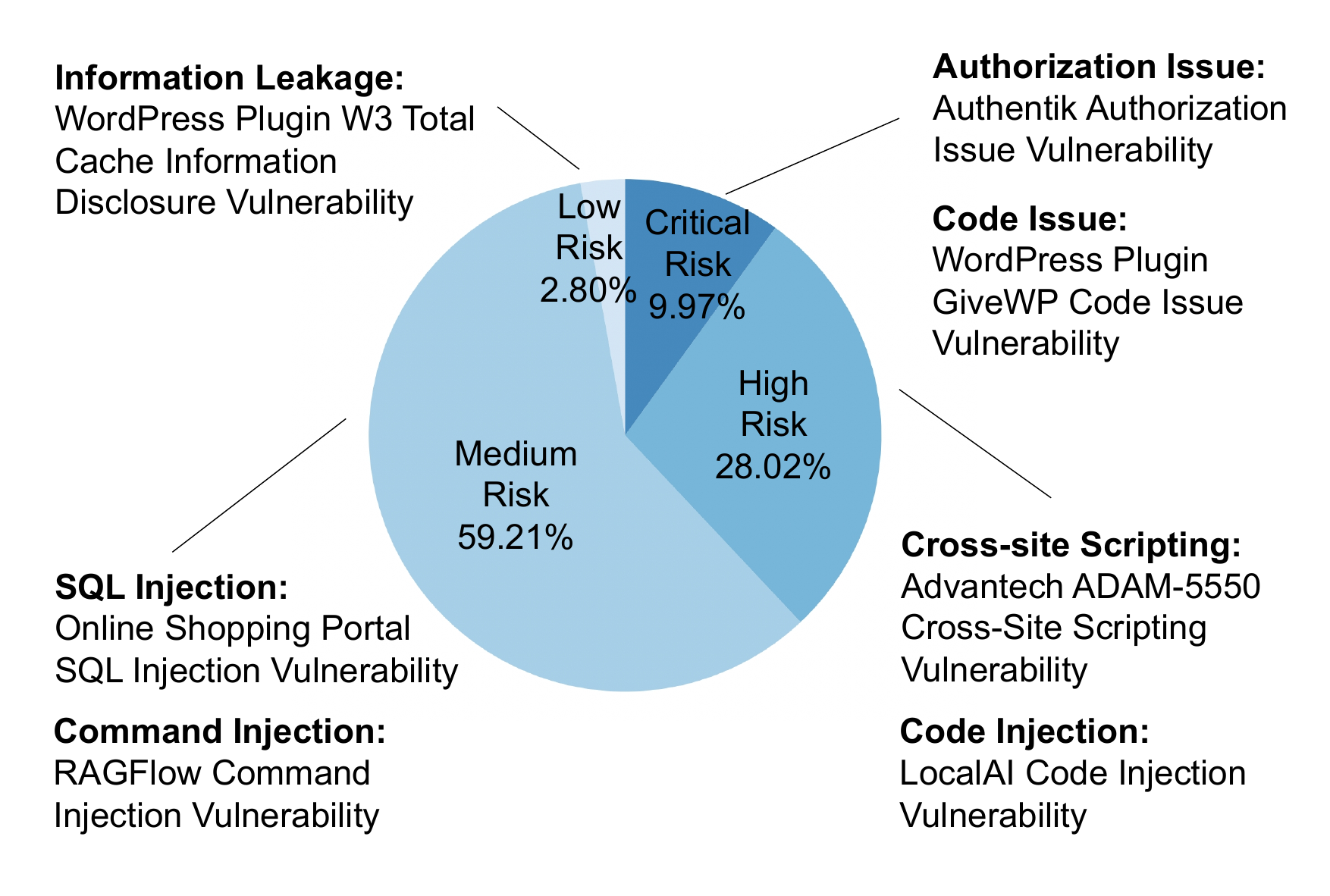}
        \label{fig:cnnvd}
    \end{minipage}
    }
    \caption{Recent situation of IDS.}
    \label{fig:motivation}
\end{figure}

IDS have been increasingly challenged to effectively deal with intrusions for decades. It is noted in Figure~\ref{fig:cve} that the number of CVE\footnote{Common Vulnerabilities and Exposures (CVE) is a security project for security information sharing and vulnerability management. CVE is a publicly accessible database where each vulnerability has a common name and a unique identifier.} records has presented an accelerating uptrend, especially in 2016, which suffered a sharp rise. After 2016, the number of CVE records stays growing at a high speed, reaching around 30,000 in 2024. Besides, according to the CNNVD\footnote{Chinese National Vulnerability Database (CNNVD) is a Chinese national database that catalogs security vulnerabilities in software and hardware products. CNNVD also provides unique identifiers and descriptions similar to CVE.} report shown in Figure~\ref{fig:cnnvd}, we can observe that almost all (i.e., 97.2\%) vulnerabilities are medium risk or above, with high and critical risk accounting for 40\% of them. The growing number of vulnerabilities and the large percentage of high-risk vulnerabilities both reveal the tough situation faced by IDS.

Nevertheless, an interesting observation from Figure~\ref{fig:cve} is that, against the number of CVE records, DL-IDS papers also started to emerge in 2016 and their amount grew year by year subsequently. We can notably find that the growth trend of DL-IDS papers is nearly the same as that of CVE records. The potential reason can be speculated as DL is an effective way for IDS to cope with their tough situation. Borrowing the strong generalizability from DL techniques, DL-IDS detection can be extended to zero-day intrusions that are almost impossible to detect with the traditional DL-IDS. Some studies \cite{r24, r30, r29} demonstrate this speculation. In their experiments, DL-IDS are all reported with an achievement of over 90\% detection accuracy while the traditional DL-IDS sometimes only have around 50\% detection accuracy.

The IDS future is \textit{not only tough but also bright} with the aid of DL - it is evident that the growth in the number of IDS papers primarily comes from those based on DL techniques. The proportion of DL-IDS papers rises from about 0\% in 2016 to a very high 65.7\% in 2024. This phenomenon reflects the great interests and visions of the cybersecurity community in DL-IDS. To date, the DL-IDS development has almost reached a decade, and thus, it is time, and also essential, to revisit how DL and IDS interact, identify emerging trends, and guide future research directions.

\subsection{Related Surveys and Our Scope}

Unfortunately, none of the related surveys in the last decade have systematically investigated DL-IDS. On one hand, some related surveys may only focus on a few parts of DL-IDS, such as log parsers \cite{r3, r4, r5}, datasets \cite{r1}, attack modeling \cite{r1, r8}, and specific DL technique type \cite{bilot2023graph}. On the other hand, while several surveys \cite{r2, r6, r7, r9, r10, r71, r121, r78, r73, r110, liao2013intrusion} involve some DL-based approaches, they did not review DL-IDS from the perspective of DL particularly.

\paragraph{Partial Investigation for DL-IDS}
The surveys \cite{r5, r3, r4, r1, r8} are the typical example papers describing only a few parts of DL-IDS. 
Among them, Adel et al. \cite{r8} mainly studied various techniques and solutions that were tailored to APT attacks, as well as discussed where to make the APT detection framework smart.
Scott et al. \cite{r4} and Tejaswini et al. \cite{r3} detailedly discussed online log parsers and their applications for anomaly detection.
Branka et al. \cite{r1} review APT datasets and their creation, along with feature engineering in attack modeling.
Zhang et al. \cite{r5} created an exhaustive taxonomy of system log parsers and empirically analyzed the critical performance and operational features of 17 open-source log parsers.
Tristan et al. \cite{bilot2023graph} focused on the applications of graph neural networks (GNNs) to IDS.
For DL-IDS, all the above surveys are obviously insufficient to advance research understanding and provide theoretical suggestions.

\paragraph{Different Perspectives from DL-IDS} 

Another type of existing surveys involved DL-IDS but studied them from the other perspectives \cite{r2, r6, r7, r9, r10, r71, r121, r78, r73, r110, liao2013intrusion, ahmad2021network}. Specifically, the surveys \cite{r110, liao2013intrusion} aim to give an elaborate image of IDS and comprehensively explain methods from signature checking to anomaly detection algorithms. Originating from log data, the survey \cite{r2} presented a detailed overview of automated log analysis for reliability engineering and introduced three tasks including anomaly detection, failure prediction, and failure diagnosis. In survey \cite{r73}, Nasir et al. explored the efficacy of swarm intelligence on IDS and highlighted the corresponding challenges in multi-objective IDS problems.

Additionally, data types inspire and contribute significantly to the related surveys, whose categories include host-based IDS (HIDS) \cite{r121, r78, r7, r9, r10} and network-based IDS (NIDS) \cite{r71, ahmad2021network}. Bridges et al. \cite{r78} focused on IDS leveraging host data for the enterprise network. Martins et al. \cite{r121} brought the HIDS concept to the Internet of Things. As a representative form of data in HIDS, the provenance graph \cite{r7, r9, r10} and its reduction techniques \cite{r6} were also extensively studied in survey literature. In NIDS, Nassar et al. \cite{r71} studied the techniques of network intrusion detection, especially those with machine learning (ML). Ahmad et al. \cite{ahmad2021network} further incorporated ML and DL into their NIDS survey and studied the downstream learning methods detailedly.

The above surveys, however, lack investigation and discussion about DL-IDS. DL techniques are only what they cover or involve, rather than the primary focus of their research.

\paragraph{Scope of Our Survey} 
Our work distinguishes the related surveys by providing a comprehensive literature review of DL-IDS. From the perspective of DL, our survey elaborates on a common workflow of DL-IDS and introduces the corresponding taxonomies of all modules within this workflow. Moreover, our survey discusses the possible challenges and research visions for DL-IDS, which include many DL-related issues that have not yet been studied by the existing surveys. 

\subsection{Contributions and Organization}

In summary, this survey makes the following contributions:

\begin{itemize}
\item Realizing that IDS has made significant progress with the aid of DL over the last decade, we present a thorough survey for DL-IDS, formalizing its definition and clarifying its location among other types of IDS.
\item We outline the common workflow for DL-IDS, consisting of the data management stage and intrusion detection stage. We further systematically illustrate the research advances in all modules of this workflow and innovatively taxonomize the papers based on DL techniques.
\item From the perspective of DL, we discuss the potential challenges and future directions for DL-IDS, especially highlighting the ones unique to DL-IDS for accommodating current researchers.
\end{itemize}

\paragraph{Survey Structure}
Section~\ref{sec:methodology} introduces the survey methodology of this work.
Section~\ref{sec:background} describes the background knowledge about DL-IDS.
Section~\ref{sec:data} and Section \ref{sec:model} elaborates the recent research trends on data management stage and intrusion detection stage, respectively.
Section~\ref{sec:benchmark} illustrates the benchmark datasets and their feature dimensions. 
Section~\ref{sec:discussion} discusses the visions and challenges for future research.
Lastly, the conclusion is presented in Section~\ref{sec:conclusion}. 

\section{Survey Methodology}
\label{sec:methodology}

\begin{figure}[t]
    \begin{minipage}[t]{0.59\linewidth}
        \centering
        \includegraphics[width=\linewidth, trim=0 25 0 0]
        {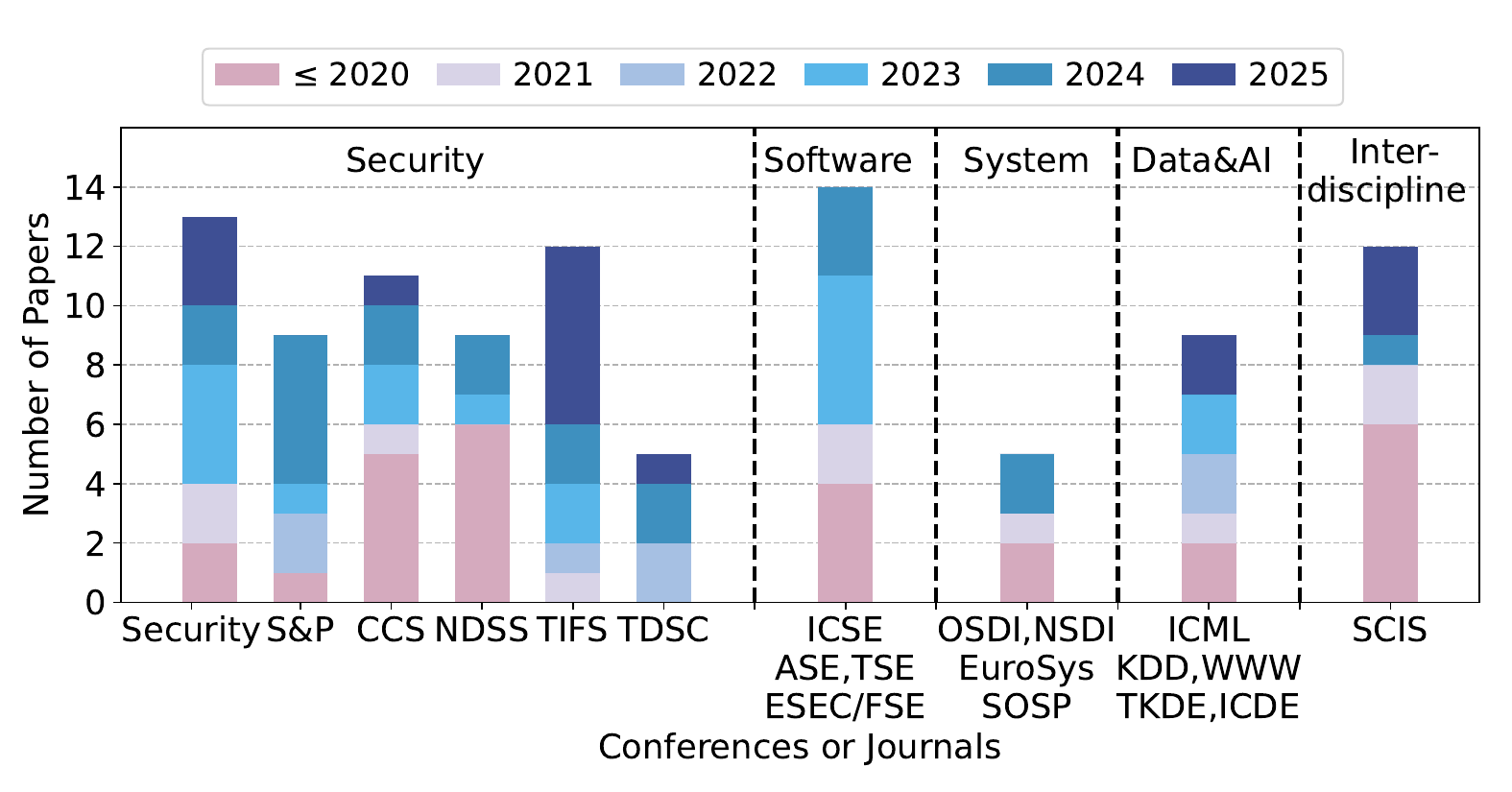}
        \caption{Source distribution of references.}
        \label{fig:bar}
    \end{minipage}
    \hspace{10pt}
    \begin{minipage}[t]{0.325\linewidth}
        \centering
        \includegraphics[width=\linewidth,trim=0 -17.5 0 0]
        {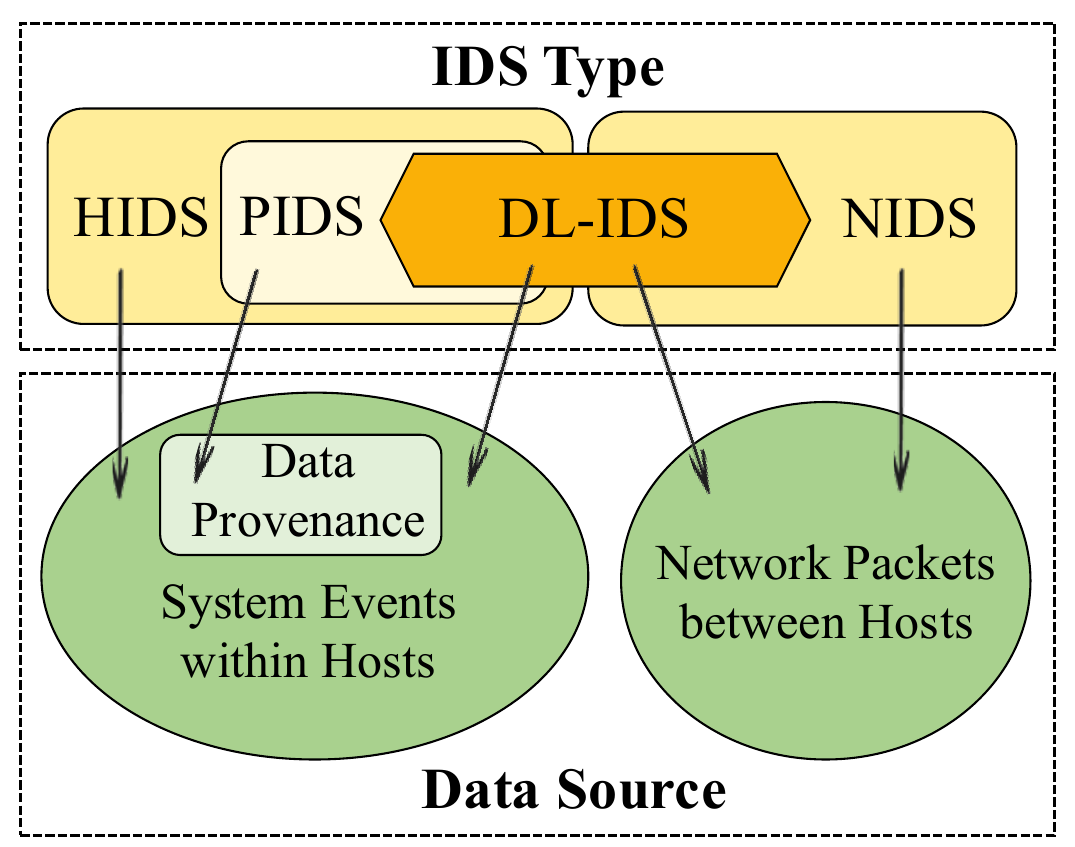}
        \caption{Types of IDS.}
        \label{fig:ids}
    \end{minipage}
    % \vspace{-20pt}
\end{figure}

To start our literature review, we selected several popular literature databases, including Web of Science \cite{webofscience}, IEEE Xplore \cite{IEEE}, and Scopus \cite{Elsevier}, as the search engine. For search keywords, we determined from generalized terms associated with DL-IDS, such as intrusion detection system, attack investigation, anomaly detection, threat detection, Advanced Persistent Threats, data provenance analysis, forensic analysis, causality analysis, log collection, log compression, log parsing, log storage, and log summarization. Then, we employed Connected Papers \cite{connectedpapers}, a visual tool that assists researchers in finding relevant academic papers, to ensure that we did not overlook the typical related literature. Since the found literature is numerous and rather generalized for the DL-IDS scope, we carefully checked their topics and prioritized only academic papers that are highly related. Finally, all these papers were filtered based on the impact factors of their published journals or academic conferences, leaving us a total of 131 papers.

We identified a few venues that have published many significant papers in the field of DL-IDS, such as Usenix Security, S\&P, CCS, NDSS, TIFS, TDSC, ICSE, ASE, ESEC/FSE, TSE, OSDI, NSDI, EuroSys, SOSP, ATC, ICML, KDD, WWW, TKDE, ICDE, and SCIS. We broadly divide them into five categories: security, software, system, data, and interdiscipline. The distribution of these papers with their published years is reported in Figure~\ref{fig:bar}.

\section{Background}
\label{sec:background}

\subsection{Intrusion Detection System}

\subsubsection{Definition of IDS}

IDS have long been a central issue in the cybersecurity community, whose research can be traced back to the 1990s \cite{bace2001intrusion} or even earlier. According to the existing literature \cite{bace2001intrusion, liao2013intrusion, r71, r72, r73, r130}, IDS can be defined progressively as follows:
\begin{definition}
\label{def:ids}
\textit{(Intrusion Detection System)}. Intrusion detection system is a software or hardware system to automate the process of \textit{intrusion detection}.
\end{definition}
\begin{definition}
\label{def:id}
\textit{(Intrusion Detection)}. Intrusion detection is the process of monitoring and analyzing the events occurring in a computer or a network for signs of \textit{intrusion}s.
\end{definition}
\begin{definition}
\label{def:i}
\textit{(Intrusion)}. Intrusion is the attempt to undermine the confidentiality, integrity, and availability of a computer or a network, or to circumvent its security facilities.
\end{definition}

\subsubsection{Types of IDS}
\label{sec:ids_types}

Generally, IDS can be further categorized into various types based on their data sources \cite{r10}. Well-known types include NIDS, HIDS, and Provenance-based IDS (PIDS). Figure~\ref{fig:ids} depicts IDS types, their data sources, and the location of DL-IDS within those IDS types.

\begin{definition}
\label{def:nids}
\textit{(NIDS)}. NIDS are IDS whose data sources are network traffic between hosts.
\end{definition}
NIDS takes network traffic between hosts as its input. It is usually deployed at the edge or key node of the network, allowing it to secure the whole computer system with limited data. Benefiting from the global perception of the whole computer system, NIDS does well in large-scale multi-host intrusions such as Distributed Denial-of-Service (DDoS) attacks. However, NIDS performs poorly in intra-host intrusions and is difficult to analyze intrusions in the form of encrypted network traffic.

\begin{definition}
\label{def:hids}
\textit{(HIDS)}. HIDS are IDS whose data sources are system events within hosts.
\end{definition}
HIDS, in contrast, uncovers intrusions through system events of individual hosts. Its data sources include file system changes, system calls, process activities, etc. HIDS can conduct comprehensive detection for a host, and is not affected by encrypted data since the decryption is also performed in the host. Nevertheless, the deployment and maintenance of HIDS is relatively difficult. HIDS should be adapted to hosts of different operating systems and runtime environments. This simultaneously introduces computation overhead for the hosts.

\begin{definition}
\label{def:pids}
\textit{(PIDS)}. PIDS are HIDS whose data sources are \textit{data provenance}.
\end{definition}
\begin{definition}
\label{def:dp}
\textit{(Data Provenance)}. Data provenance refers to the origin and the processes that an event has undergone from its creation to its current state.
\end{definition}

PIDS is a subtype of HIDS, particularly referring to HIDS that utilizes data provenance as its data source. Due to analysis in the intact trail of events, PIDS is proven effective in coping with advanced attacks \cite{r10}. By performing causality analysis on data provenance, PIDS can significantly reduce false alarms. Yet, data provenance is very expensive to obtain, requiring complicated technical tools for monitoring operating systems, network protocols, and applications. 

\begin{definition}
\textit{(DL-IDS.)} DL-IDS are IDS that utilize DL techniques to detect intrusions, whose data sources can be network traffic between hosts, system events within hosts, or their combination.
\end{definition}

Unlike the other types of IDS such as NIDS and HIDS are categorized by their data sources, DL-IDS is defined by the techniques used in intrusion detection. As shown in Figure~\ref{fig:ids}, the data source of DL-IDS can be network traffic, system events, or both. Taking advantage of the generalizability of DL techniques, DL-IDS is allowed to handle zero-day attacks precisely and thus become extremely interested in the cybersecurity community recently.

\begin{figure}[t]
    \centering
    \includegraphics[width=\linewidth,trim=0 0 0 0]{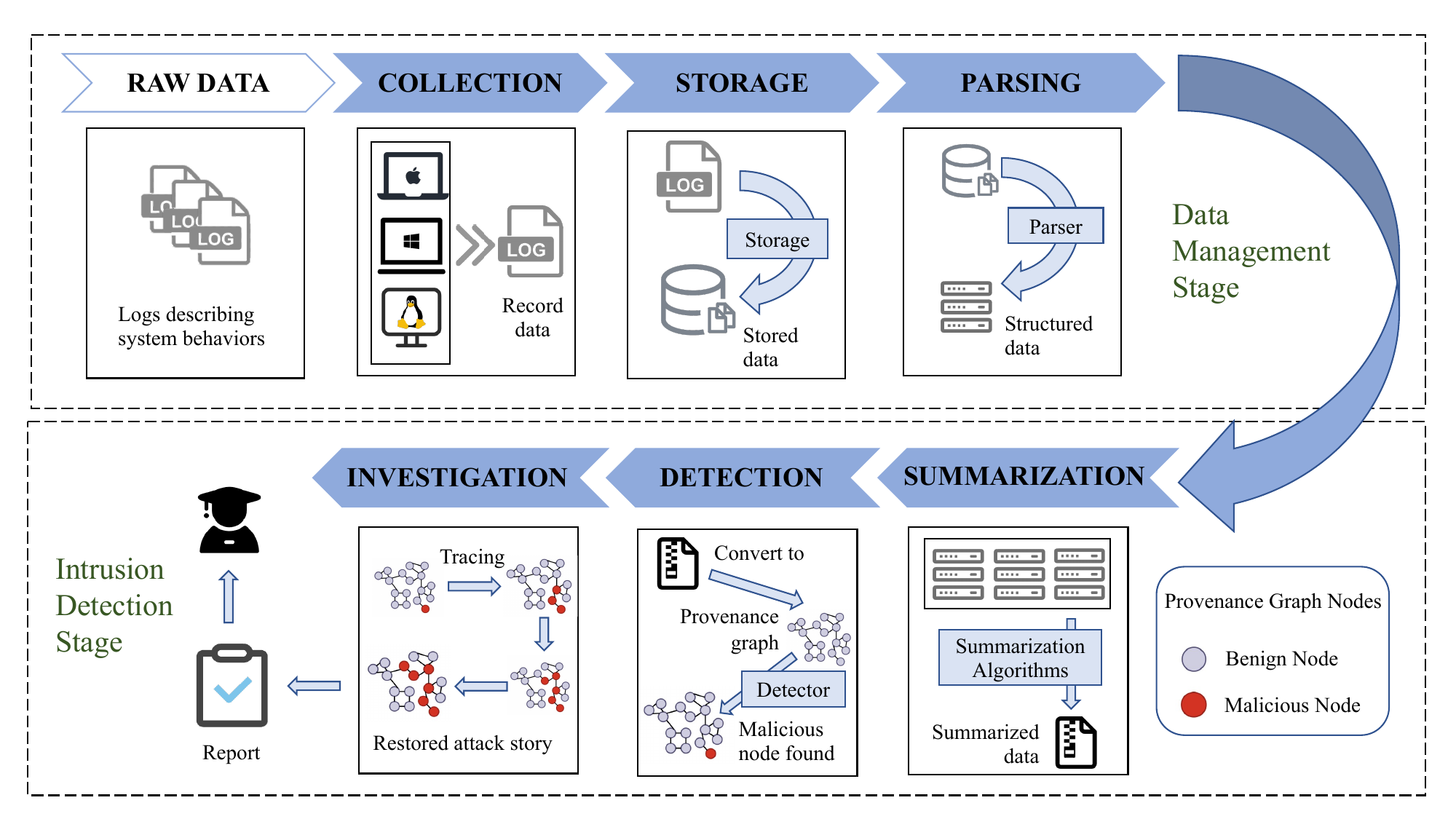}
    \caption{Common workflow of DL-IDS.}
    % \vspace{-20pt}
    \label{fig:workflow}
\end{figure}

\subsection{Common Workflow}
\label{sec:workflow}

Figure~\ref{fig:workflow} depicts the common workflow of DL-IDS. It usually consists of 7 steps: raw data, collection, storage, parsing, summarization, detection, and investigation, which are explained as follows:

\begin{itemize}
\item \textbf{Raw Data} is unprocessed data for uncovering attack details or benign system behaviors. The raw data analyzed by cyber experts commonly include network traffic and audit logs.
\item \textbf{Collection} indicates data collection tools for different systems, such as cloud and cross-platforms, which gather valuable raw data to describe important system behavior scenarios.
\item \textbf{Storage} involves storage and search engines to manage large amounts of collected log data. Log data is labeled with indexes for efficient retrieval. 
\item \textbf{Parsing} is the act of analyzing the stored logs and other useful data. It extracts and organizes the underlying information within the data for subsequent processing.
\item \textbf{Summarization} refers to the operation of summarizing large volumes of parsed data based on its semantics. This reduces storage costs while preserving critical events.
\item \textbf{Detection} is the process of using detection tools such as models and algorithms to detect anomalies in analyzed data to determine whether the data contains intrusions.
\item \textbf{Investigation} is the further process of Detection. It reconstructs the entire attack scenarios from the detected malicious data by analyzing the causal relationship between them.
\end{itemize}

\noindent Note that DL-IDS can also be performed in other step orders by skipping some of the steps. For example, log data can be first parsed before storage \cite{r89}. Attack investigation can be directly conducted without detection of intrusions \cite{r69}. This survey is organized by the common workflow.

\section{Data Management}
\label{sec:data}

This section elaborates on the data management stage of DL-IDS, including data collection (Section~\ref{sec:collection}), log storage (Section~\ref{sec:storage}), and log parsing (Section~\ref{sec:parsing}).

\subsection{Data Collection}
\label{sec:collection}

The first step of DL-IDS is to collect useful data from raw data. Raw data indicates records that document events, activities, and operations that occur within a system, application, or network (a.k.a., logs), represented by audit logs or application logs within hosts, or network traffic between hosts. By collecting useful logs, DL-IDS is allowed to monitor the health condition and operational status of information systems \cite{r5, r139}. Common attributes of logs include timestamp, event type, subject, object, description, etc.

On different platforms, logs possess different formats and organizational structures \cite{r5, r10, r78, r9}. To counter this, researchers have created various log collection tools specialized for various systems. For example, in Windows systems, Event Viewer is employed to manage system logs. Yet in Linux systems, log files are usually saved in the /var/log/ directory. The classification of data collection tools is shown in Table~\ref{tab:collection}, including Windows, Linux, Cloud, and Cross platforms.

\begin{table}[t]
\centering
  \caption{Log collection tools on different platforms.}
  % \vspace{-8pt}
  % \setlength\tabcolsep{8pt}
  \label{tab:collection}
  \footnotesize
  \begin{tabular}{ccl}
    \toprule
    \textbf{Platform Type}&\textbf{\makecell{Tool}}&  \hspace{2cm}\textbf{Description}\\
    \midrule
    \multirow{2}{*}{Windows platform}& ETW \cite{r11} & Providing developers comprehensive event tracing ability\\
    & Panorama \cite{r12} & Hardware-level and OS-aware dynamic taint tracking\\
    \hline
    \multirow{8}{*}{
    Linux platform} & auditd \cite{auditd} & Native tools supported by the Linux kernel \\
    & sysdig \cite{kilisysdig} & Focusing on runtime monitoring and fault troubleshooting \\
    & CamFlow \cite{r15} & Self-contained, easily maintainable implementation \\& Tracee \cite{tracee2022runtime} & Exposing system information as events based on eBPF \\
    & DataTracker \cite{r13} & Monitoring unmodified binaries without their source codes \\
    & Inspector \cite{r14} & Parallel provenance library that is POSIX-compliant \\& AutoLog \cite{r16} & Analyzing programs so no need to run them\\
    & \textit{e}Audit \cite{sekar2024eaudit} & Fast, scalable and easily deployable data collection tools\\
    \hline
    \multirow{3}{*}{Cloud platform} &K8S tools \cite{r19, r20} &
    Adapting to cloud scenarios to meet enterprise needs\\
    & saBPF \cite{lim2021secure} & An extension tool of eBPF for containers in cloud computing\\
    & ISDC \cite{mirnajafizadeh2024enhancing} & Eliminating overheads on in-network resources\\
    \hline
    \multirow{2}{*}{Cross platform} &DTrace \cite{r17} & Real-time tracing framework that supports many platforms \\
    & SPADE \cite{r18} & Novel provenance kernel for cross-platform logging\\
  \bottomrule
  % \vspace{-15pt}
\end{tabular}
\end{table}

\subsubsection{Windows Platform Tools} 

Event Tracing for Windows (ETW) \cite{r11} is a powerful event tracing mechanism provided by Microsoft. It consists of three components: providers, controllers, and consumers. ETW instruments applications to provide kernel event logging and allows developers to start and stop event tracing sessions momentarily.
Panorama \cite{r12} exploits hardware-level and OS-aware dynamic taint tracking to collect logs. Moreover, it develops a series of automated tests to detect malware based on several kinds of anomalous behaviors.

\subsubsection{Linux Platform Tools} 

auditd \cite{auditd} is a native collection tool supported by the Linux kernel, which is responsible for writing audit logs to disk and monitoring a variety of auditable events such as system calls, file accesses, and modifications.
sysdig \cite{kilisysdig} relies on the kernel module to achieve monitoring and data collection of the system. sysdig focuses on system runtime monitoring and fault troubleshooting, which is also widely used in containers and cloud-native environments.\
CamFlow \cite{r15} designs a self-contained, easily maintainable implementation of whole-system provenance based on Linux Security Module, NetFilter, and other kernel facilities. Furthermore, it provides a mechanism to adapt the captured data provenance to applications and can be integrated across distributed systems.
Tracee \cite{tracee2022runtime} takes advantage of the extended Berkeley Packet Filter (eBPF) framework to observe systems efficiently. It uses eBPF to tap into systems and expose that information as events. 
DataTracker \cite{r13} is an open-source data provenance collection tool using dynamic instrumentation. It is able to identify data provenance relations of unmodified binaries without access to or knowledge of the source codes.
Inspector \cite{r14} is a Portable Operating System Interface (POSIX)-compliant data provenance library for shared-memory multi-threaded applications. It is implemented as a parallel provenance algorithm on a concurrent provenance graph.
AutoLog \cite{r16} generates runtime log sequences by analyzing source codes and does not need to execute any programs. It can efficiently produce log datasets (e.g., over 10,000 messages/min on Java projects) and has the flexibility to adapt to several scenarios.
\textit{e}Audit \cite{sekar2024eaudit} is a scalable and easily deployable data collection tools. \textit{e}Audit 
relies on the eBPF framework built into recent Linux versions, making it work out of the box on most of the Linux distributions.

\subsubsection{Cloud Platform Tools} 

Although some collection tools in Windows and Linux platforms such as auditd \cite{auditd}, sysdig \cite{kilisysdig}, and Tracee \cite{tracee2022runtime} can be applied in cloud computing environment, cloud-native scenarios introduce different challenges compared with Windows or Linux platforms. First, there are many different types of components such as containers, microservices, and Kubernetes (K8S) clusters in cloud platforms, each of which generates its own logs with varying formats and contents. Additionally, components are basically characterized by dynamic expansion and contraction, making it hard to capture complete log data. 
To address them, Chen et al. \cite{r19} design a cloud log collection architecture on the basis of K8S, which is a central platform based on cloud-native technology.
Josef et al. \cite{r20} propose a log collection and analysis tool operated as Software as a Service (SaaS) in the cloud environment in K8S technology, aiming to provide comprehensive logs across all microservices.
saBPF \cite{lim2021secure} is an extension tool of eBPF, aiming to deploy fully-configurable, high-fidelity, system-level audit mechanisms at the granularity of containers. saBPF is further developed with proof-of-concept IDS and access control mechanism to demonstrate its practicability.
ISDC \cite{mirnajafizadeh2024enhancing} is designed to eliminate the bottleneck between network infrastructure (where data is generated) and security application servers (where data is consumed), which prioritizes specific flows to effectively optimize resource consumption.

\subsubsection{Cross-platform Tools} 

To effectively detect intrusions, an intuitive idea is to incorporate log data from various platforms to obtain a global view of the running system.
DTrace \cite{r17} is a real-time dynamic tracing framework for troubleshooting kernel and application problems on production systems. It supports many platforms, including 	Linux, Windows, Solaris, macOS, FreeBSD, NetBSD, etc.
Support for Provenance Auditing in Distributed Environments (SPADE) \cite{r18} develops a novel provenance kernel that mediates between the producers and consumers of provenance information, and handles the persistent storage of records. It supports heterogeneous aggregating for system-level data provenance for data analysis across multiple platforms. 

\subsection{Log Storage}
\label{sec:storage}

The subsequent step of log collection is to store these logs \cite{r129, altinisik2023provg}. We will introduce two essential components for data storage: log storage systems and compression algorithms for these systems.

\subsubsection{Log Storage Systems}
\label{sec:storage-systems}

The two most commonly used log storage systems are ELK \cite{r79} and Loki \cite{r80}. ELK is a powerful log management solution consisting of three open-source software components: Elasticsearch \cite{elasticsearch}, Logstash \cite{logstash}, and Kibana \cite{kibana}. Elasticsearch \cite{elasticsearch} is the leading distributed, RESTful search and analytics data engine designed with speed and scalability. Logstash \cite{logstash} is a server-side data preprocessing pipeline to collect and integrate data from multiple sources. Kibana \cite{kibana} is a data analytics and visualization platform at both speed and scale. ELK is powerful enough to be applied in enterprise scenarios, however, its performance comes at a price. ELK sacrifices ease of configuration and installation, and may simultaneously introduce severe runtime overhead for its hosts. In contrast, Loki \cite{r80} is a lightweight logging system with low resource overhead developed by Grafana Labs. It is designed with simple operations and efficient storage. Instead of indexing everything of data like ELK does, Loki mainly creates indices grounded in log labels. Moreover, Loki is well suited for open-source monitoring and visualization tools such as Prometheus \cite{prometheus} and Grafana \cite{grafana}. Integrating these two tools enables Loki to construct a complete monitoring and log analysis platform for information systems.

\subsubsection{Log Compression Algorithms}
\label{sec:compression}
Logs are generated quickly and require significant memory usage. For example, it is measured that a browser can produce about 10 GB of log data each day \cite{r129}. Such oversize data should be compressed before storage. Log compression algorithms can be categorized into two types: general-purpose algorithms and those specifically adapted to log data.

\paragraph{General Compression Algorithms}

\begin{table}[t]
    \centering
    \footnotesize
    \caption{Well-acknowledged general compression algorithms for log data.}
    % \vspace{-5pt}
    \label{tab:compression}
    \begin{tabular}{cl}
    \toprule
    \textbf{Type} & \makecell{\textbf{Well-acknoledged compression algorithm}}\\
    \midrule
    Dictionary-based &  LZ77 in \textit{gzip} \cite{gzip}, LZMA in \textit{7zip\_lzma} \cite{lzma}, and LZSS in \textit{quickLZ} \cite{quicklz}\\
    Sorting-based & BWT in \textit{bzip2} \cite{bzip2} andST in \textit{szip} \cite{szip}\\
    Statistical-based & PPMD in \textit{7zip\_ppmd} and DMC in \textit{ocamyd} \cite{ocamyd}\\
    \bottomrule
    \end{tabular}
    % \vspace{-10pt}
\end{table}

General compression algorithms refer to algorithms to reduce the size of data (e.g., log data) by handling token-level or byte-level duplicates in the data. General compression algorithms can be classified into three categories based on their principles \cite{r86}:
\begin{itemize}
    \item Dictionary-based Compression: It records repeated data as keys and replaces these data with their corresponding keys.
    \item Sorting-based Compression: It sorts data to enable strategies that require ordering features.
    \item Statistical-based Compression: It exploits statistical techniques to learn and predict the possible next token for existing tokens. The data is thus compressed as a statistical model.
\end{itemize}
Table~\ref{tab:compression} presents representative algorithms of the above three types. Due to the indeterminacy of statistical techniques, statistical-based compression algorithms may introduce losses in compression. Yet the other two types of algorithms are generally lossless. By validating 9 log files and 2 natural language files, a study \cite{r86} shows that some general compression algorithms can achieve high compression ratios for log data and log data is even easier to compress than natural language data.

\paragraph{Tailored Compression Algorithms}

Different from natural language data, log data usually has specific structures and formal expressions that help further compression. Yao et al. \cite{r87} propose LogBlock, which obtains small log blocks before compression and then uses a generic compressor to compress logs. Liu et al. \cite{r89} propose Logzip, which employs clustering algorithms to iteratively extract templates from raw logs and then obtain coherent intermediate representations for compressing logs. Rodrigues et al. \cite{r92} propose the lossless compression tool CLP, aiming to quickly retrieve log data while meeting compression requirements. CLP proposes to combine domain-specific compression and search with a generic lightweight compression algorithm. Li et al. \cite{r88} conduct empirical research on log data and propose LogShrink to overcome their observed limitations by leveraging the commonality and variability of log data. LogBlock \cite{r87} is designed to help existing jobs perform better. It reduces duplicate logs by preprocessing log headers and rearranging log contents, thereby improving the compression ratio of log files. LogReducer \cite{yu2023logreducer} is a framework that combines log hotspot identification and online dynamic log filtering. Its non-intrusive design significantly reduces log storage and runtime overhead. $\mu$Slope \cite{wang2024muslope} is a compression and search method for semi-structured log data. It achieves efficient storage and query performance through data segmentation, pattern extraction, and index-free design. Denum \cite{yu2024unlocking} significantly improves log compression rates by optimizing the compression of digital tokens in logs. It is an efficient log compression tool suitable for scenarios where you need to save storage space or transmission bandwidth.

\subsection{Log Parsing}
\label{sec:parsing}

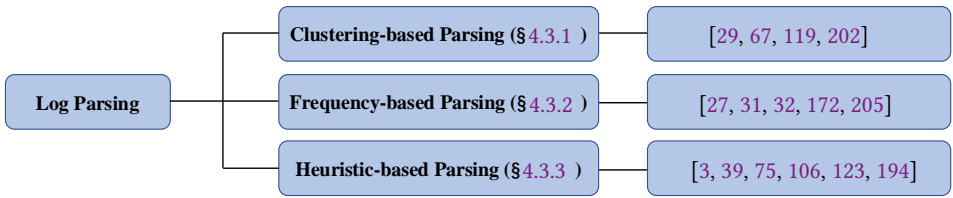
\begin{figure}[t]
    \centering
    \begin{overpic}[scale=0.45, tics=3, width=0.925\linewidth, trim=0 0 0 0]{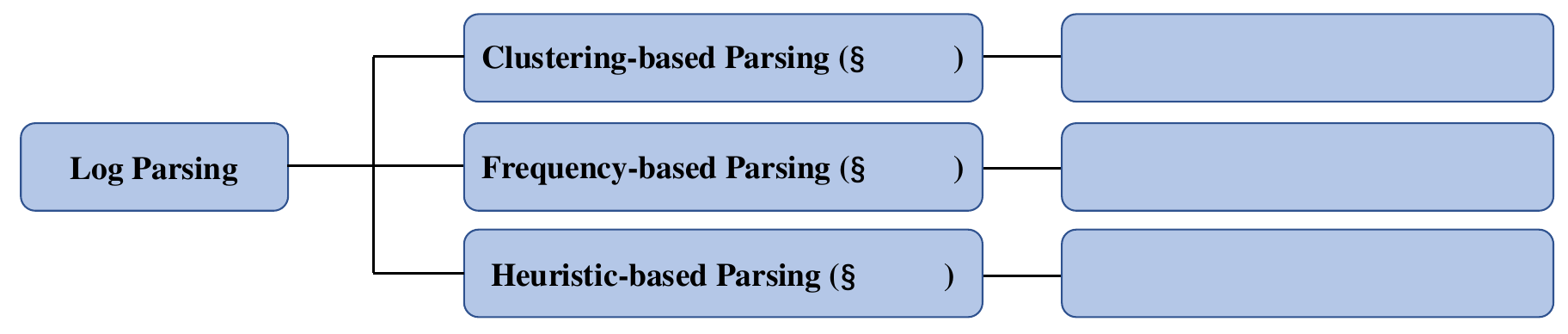} 
        \small
        \put(55.4,16.6){\ref{sec:clutering}}
        \put(55.5,9.6)
        {\ref{sec:frequency}}
        \put(54.8,2.8)
        {\ref{sec:heuristic}}
        \put(74,16.6)
        {\cite{r94, hamooni2016logmine, lin2016log, coustie2020meting}}
        \put(73,9.6)
        {\cite{r96, r97, r98, r100, chu2021prefix}}
        \put(72.2,2.8)
        {\cite{r101, r102, r103, du2016spell, agrawal2019logan, vervaet2021ustep}}
    \end{overpic}
    \caption{Taxonomy of data parsing.}
    % \vspace{0pt}
    \label{fig:parsing}
\end{figure}

Log data often originates from multiple different devices such as terminals, sensors, and network devices. To analyze it, log parsers are employed to format them into structured and unified ones. Log parsing is usually executed by data classification and template extraction. Data classification is to classify log data into several groups. Each group constitutes a template for extracting features from log data and constructing the structured logs. As shown in Figure~\ref{fig:parsing}, the existing log parsers can be taxonomized into 3 categories: clustering-based, pattern-based, and heuristic-based parsers.

\subsubsection{Clustering-based Parsing.} 
\label{sec:clutering}

Clustering-based parsers classify data using clustering algorithms for log parsing. Xiao et al. \cite{r94} propose LPV, which employs a hierarchical clustering algorithm to incrementally group logs based on Euclidean distance.
Hamooni et al. \cite{hamooni2016logmine} present a rapid log pattern recognition approach named LogMine. It is implemented in the map-reduce framework for distributed platforms to process millions of log messages in seconds.
LogCluster \cite{lin2016log} reduces the number of logs that need to be manually checked and improves the accuracy of problem identification through log clustering and the use of knowledge bases.
METING \cite{coustie2020meting} provides a robust and efficient log parsing method through frequent n-gram mining and flexible log grouping strategy, which can effectively process various types of log data.

\subsubsection{Frequency-based Parsing.} 
\label{sec:frequency}
Frequency-based parsers discover patterns that exceed the frequency threshold and employ the mined patterns to parse logs. Sedki et al. \cite{r96} propose the log parsing tool ULP, which combines string matching and local frequency analysis to efficiently parse large log files. Dai et al. \cite{r97} propose Logram, which utilizes an n-gram dictionary for log parsing. For n-grams with a frequency below the threshold, Logram recursively converts to (n-1)-grams until a list of uncommon 2-grams is obtained. To mitigate the parameter sensitivity issue in log parsers, Dai et al. \cite{r98} further proposed an entropy-based log parser PILAR, which balances parsing accuracy and efficiency. Xu et al. \cite{r100} propose a hybrid log parsing model called Hue, which performs parsing through user-adaptive methods. Prefix-Graph \cite{chu2021prefix} is an efficient, adaptive, and universal log parsing method that can stably extract log templates without relying on domain knowledge and manual parameter tuning.

\subsubsection{Heuristic-based Parsing.} 
\label{sec:heuristic}

Heuristic-based parsers rely on empirical knowledge to classify log data. He et al. \cite{r101} propose the online log parsing method Drain, which employs a depth-fixed parsing tree to group the original logs and encodes them using specially designed parsing rules. Le et al. \cite{r102} propose to use a hint-based few-sample learning algorithm, LogPPT, to capture log template patterns. Utilizing new prompt tuning methods and an adaptive random sampling algorithm, LogPPT performs well on multiple public datasets. Liu et al. \cite{r103} propose the UniParser parser to address the issue of difficult processing of heterogeneous logs, using the Token Encoder and Context Encoder modules to learn log context features. Spell \cite{du2016spell} is an efficient streaming log parsing method that can dynamically extract log patterns in online processing and significantly improve processing efficiency through pre-filtering steps. Logan \cite{agrawal2019logan} achieves efficient and scalable log parsing through distributed processing, LCS matching, dynamic matching tolerance, and periodic merging. USTEP \cite{vervaet2021ustep} is an online log parsing method based on an evolutionary tree structure that can discover and encode new parsing rules. It achieves constant parsing time and can efficiently parse raw log messages in a streaming manner.

\section{Intrusion Detection}
\label{sec:model}

The intrusion detection stage uncovers intrusions relying on the \textit{semantic-level} information. This section classifies and summarizes the mainstream graph summarization (Section~\ref{sec:summarization}), attack detection (Section~\ref{sec:detection}), and attack investigation (Section~\ref{sec:investigation}).

\subsection{Graph Summarization}
\label{sec:summarization}

It is illustrated that stealthy malware will inevitably interact with the underlying OS and be captured by provenance monitoring systems \cite{r26}, which is the reason why PIDS (a form of DL-IDS) has worked and flourished recently. Log data generated from provenance monitoring systems is referred to as data provenance as mentioned. Offering advantages in high precision, data provenance sacrifices memory performance to record all trails of events from their creations to their current states, even some of which are trivial. Unlike network traffic and application logs, data provenance is fine-grained, detailed, and rich in semantics. As a result, the token-level or byte-level log storage systems (Section~\ref{sec:storage-systems}) and log compression algorithms (Section~\ref{sec:compression}) are insufficient to handle the memory efficiency of data provenance due to the absence of semantic-level information.

To this end, graph summarization is investigated to further reduce the size of log data semantically. In graph summarization, data provenance is transformed into a provenance graph, of which the causal relations are utilized to build the semantic understanding of system activities. Referring to the definition of data provenance (Definition~\ref{def:dp}), provenance graph is defined as follows:
\begin{definition}
\label{def:pg}
\textit{(Provenance Graph)}. Provenance graph is a representation of a collection of data provenance with causal relations. It is a directed acyclic graph $G=<V,E>$ where nodes $V$ are system entities and edges $E$ are system events.
\end{definition}
Provenance graphs allow graph summarization approaches to reduce the size of log data by confidently removing irrelevant events, aggregating similar events, gathering similar execution entities, etc. This categorizes them as a type of lossy reduction, yet the aforementioned log storage and compression are usually lossless (except for statistical-based log compression). We note that some surveys (e.g., \cite{r10,r6}) may interchangeably use graph summarization and log compression to identify the approaches that reduce the size of log data. In this work, we explicitly distinguish them and refer to the lossless reduction as compression and the opposite one as summarization. Table~\ref{tab:summarization} presents the overview of graph summarization approaches. We classify them into two categories: offline graph summarization and online graph summarization.

\begin{table}[t]
  \centering
  \caption{Overview of graph summarization approaches.}
  \setlength\tabcolsep{10pt}
  \label{tab:summarization}
  \footnotesize
  % \vspace{-6pt}
  \begin{tabular}{ccccc}
    \toprule
    \textbf{Mode}&\textbf{Approach}&\textbf{Release}&\textbf{Baseline}&\textbf{Requirement}\\
    \midrule
    \multirow{6}{*}{Offline} & ProvCompress \cite{r132} & 2011 & No Summarization & None\\
    & BEEP \cite{lee2013high} & 2013 & No Summarization & Instrumentation\\
    & LogGC \cite{r133} & 2013 & BEEP + No Summarization & Instrumentation\\
    & CPR + PCAR \cite{r134} & 2016 & No Summarization & None\\
    & FD + SD \cite{r135} & 2018 & CPR + PCAR & None\\
    & LogApprox \cite{michael2020forensic} & 2020 & GC + CPR + DPR & None \\
    & TeRed \cite{li2025tered} & 2025 & \tiny{LogGC + CPR + PCAR + F-DPR + NodeMerge} & None\\
    \hline
    \multirow{7}{*}{
    Online} & ProTracer \cite{r136} & 2016 & BEEP + No Summarization & Instrumentation \\
    & NodeMerge \cite{r137} & 2018 & No Summarization & None\\ 
    & Winnower \cite{r138} & 2018 & No Summarization & None\\
    & GS + SS \cite{zhu2021general} & 2021 & FD + SD & None \\
    & SEAL \cite{fei2021seal} & 2021 & FD & None \\
    & FAuST \cite{inam2022faust} & 2022 & CPR + DPR & None \\
    & AudiTrim \cite{sun2024auditrim} & 2024 & CPR + GS + F-DPR & None \\
  \bottomrule
\end{tabular}
% \vspace{-10pt}
\end{table}

\subsubsection{Offline Graph Summarization}

Offline graph summarization requires historical log data to provide global knowledge, which extracts log data from persistent storage, summarizes the data, and pushes back the summarized data to the persistent storage.
In 2011, Xie et al. \cite{r132} take inspiration from web graphs to summarize provenance graphs. They argue that provenance graphs have similar organizational structure and characteristics to web graphs, such as locality, similarity, and consecutiveness.
BEEP \cite{lee2013high} is developed based on the fact that a long-running execution can be partitioned into individual units. BEEP reverse engineers application binaries and instructions to perform selective logging for unit boundaries and unit dependencies.
LogGC \cite{r133} is a summarized audit log system that can be invoked at any time during the system execution.
Xu et al. \cite{r134} propose an aggregation algorithm PCR that preserves event dependencies during log data reduction. They further propose an algorithm named PCAR that utilizes domain knowledge to conduct graph summarization.
Hossain et al. \cite{r135} propose two dependency-preserving graph summarization approaches, FD and SD. FD is allowed to keep backward and forward forensic analysis results. SD preserves the results of common forensic analysis, which runs backward to find the entry points of intrusions and then runs forward from these points to unveil their impacts.
LogApprox \cite{michael2020forensic} aims to summarize the most space-intensive events found in logs, namely file I/O activity, which can account for up to 90\% of the log content.
TeRed \cite{li2025tered} employs unit tests to learn the system’s
normal behavior patterns for reducing provenance graphs, allowing it not to impact attack detection and investigation.

\subsubsection{Online Graph Summarization}

Online graph summarization performs real-time summarization for continually coming provenance graphs, rather than dealing with a static provenance graph.
ProTracer \cite{r136} alternates between system event logging and unit-level taint propagation. It has a lightweight kernel module and user space daemon for concurrent, out-of-order event processing.
NodeMerge \cite{r137} is a template-based graph summarization system for online event storage. It can directly work on the system-dependent provenance streams and compress data provenance via read-only file access patterns.
Winnower \cite{r138} is an extensible audit-based cluster monitoring system. For tasks replicated across nodes in distributed applications, it can define a model over audit logs to concisely summarize the behaviors of multiple nodes, thus eliminating the necessity of transmitting redundant audit records to the central monitoring node. 
The approach proposed by Zhu et al. \cite{zhu2021general} includes two real-time graph summarization strategies. The first strategy maintains global semantics, which identifies and removes redundant events that do not affect global dependencies. The second strategy is based on suspicious semantics. 
SEAL \cite{fei2021seal} is a novel graph summarization approach for causal analysis. Based on information-theoretic observations of system event data, it achieves lossless compression and supports real-time historical event retrieval. 
FAuST \cite{inam2022faust} is a logging daemon that performs transparent and modular graph summarization directly on system endpoints. FAuST consists of modular parsers that parse different audit log formats to create a unified in-memory provenance graph representation.
AudiTrim \cite{sun2024auditrim} is an efficient graph summarization approach that reduces log sizes without impacting user experiences, which allows adaptable deployment on different operating systems.

\subsection{Attack Detection}
\label{sec:detection}

Attack detection is located at the central position of DL-IDS. The objective of attack detection is to accurately identify malicious system events in log data while minimizing false alarms of normal system behaviors. Based on the types of log data, we categorize the attack detection approaches into audit log-based, application log-based, network traffic-based, and hybrid log-based detectors. 

The overview and taxonomy of attack detection approaches are presented in Table~\ref{tab:detection}. We note that recent years have also published many other academic papers for attack detection \cite{r22, r23, r66, r70, r126, r127, r128, wang2023tesec, hassan2020tactical}. Yet these papers are slightly related to DL-IDS, which are thus excluded in our survey for conciseness.

\begin{table}[H]
\centering
  \caption{Overview and taxonomy of attack detection approaches.}
  % \vspace{-5pt}
  \setlength\tabcolsep{4.3pt}
  \label{tab:detection}
  \footnotesize
  \begin{tabular}{ccccccc}
  
    \toprule
    \textbf{\makecell{Data\\Type}}&
    \textbf{Taxonomy}&
    \textbf{Approach}&
    \textbf{\makecell{Release\\Time}}&
    \textbf{\makecell{Base\\Model}}&
    \textbf{\makecell{Detection\\Style}}&
    \textbf{\makecell{Detection\\Granularity}}\\
    \midrule
    
    & \multirow{3}{*}{Traditional} & StreamSpot \cite{r111} & 2018 & K-Medoids & Online & Subgraph \\
    & \multirow{3}{*}{Learning} & Unicorn \cite{r25} & 2020 & K-Medoids & Online & Node, Subgraph\\
    & & DistDet \cite{dong2023distdet} & 2023 & HST & Online & Subgraph\\
    & & Velox \cite{bilot2025sometimes} & 2025 & FCN & Online & Node \\
    \cline{2-7}
    \multirow{2}{*}{Audit} & & ShadeWatcher \cite{r29} & 2022 & TransR & Offline & Node\\
    \multirow{2}{*}{Log} &  & threaTrace \cite{r30} & 2022 & GraphSAGE & Online & Node\\
    & \multirow{2}{*}{Graph} & ProGrapher \cite{r24} & 2023 & graph2vec & Online & Subgraph\\
    & \multirow{2}{*}{Neural} & MAGIC \cite{r31} & 2024 & GAT & Online & Node, Subgraph\\
    & \multirow{2}{*}{Network} & Flash \cite{r28} & 2024 & GraphSAGE & Online & Node\\
    & & R-caid \cite{goyal2024r} & 2024 & GNN & Offline & Node\\
    & & Argus \cite{xu2024understanding} & 2024 & MPNN, GRU & - & Node\\
    & & TAPAS \cite{zhang2025tapas} & 2025 & LSTM-GRU & Online & Task \\
    \hline

    &  \multirow{2}{*}{Traditional}  & Wei et al. \cite{xu2009detecting} & 2009 & PCA, TF-IDF & - & Log Entry\\
    & \multirow{2}{*}{Learning}  & Bodik et al. \cite{bodik2010fingerprinting} & 2010 & Logistic Regression & Online & Log Entry\\
    & & AMOD \cite{dong2018adaptive} & 2018 & SVM HYBRID & Online & Log Entry\\
    \cline{2-7}
    & & DeepLog \cite{r34} & 2017 & LSTM & Online & Log Entry\\
    & & LogRobust \cite{zhang2019robust} & 2019 & Attention LSTM & - & Log Entry\\
    Application &  &LogAnomaly \cite{r39} & 2019 & template2vec, LSTM & Online & Log Entry\\
    Log & \multirow{3}{*}{Sequence} & LogC \cite{r38} & 2020 & LSTM & Online & Log Entry\\
    & \multirow{3}{*}{Neural} & NeuralLog \cite{r44} & 2021 & BERT & - & Log Entry\\
    & \multirow{3}{*}{Network} & PLELog \cite{yang2021semi} & 2021 & Attention GRU & Online & Log Entry\\
    & & SpikeLog \cite{qi2023spikelog} & 2023 & DSNN & - & Log Entry\\
    & & LogCraft \cite{zhang2024end} & 2024 & Meta Learning & - & Log Entry\\
    & & Tweezers\cite{cui2024tweezers} & 2024 & GATv2, BERTweet & Online & Log Entry\\
    & & LogSer \cite{chai2024log} & 2024 & BERT & Online & Log Entry\\
    & & LogDLR\cite{zhou2025logdlr} & 2025 & Transformer, SBERT & Online & Log Entry\\

    \hline

    \multirow{21}{*}{Traffic} 
    & & NetPro \cite{li2017netpro} & 2017 & Merkle Hash Tree & Online & Route \\
    \multirow{21}{*}{Log} & & CATH \cite{guo2019cath} & 2019 & Cusp Model & Online & Flow \\
    & \multirow{2}{*}{Traditional} & Whisper \cite{fu2021realtime} & 2021 & K-Means & - & Host\\
    & \multirow{2}{*}{Learning} & SigML++ \cite{r52} & 2023 & ANN & -& Encrypted Log\\ % 4
    & & OADSD \cite{zhang2023real} & 2023 & Isolation Forest & Online &Packet\\
    & & LtRFT \cite{tang2023ltrft} & 2023 & LambdaMART & Offline & Packet\\
    & & AGC \cite{wu2025intrusion} & 2025 & Clustering & - & Packet\\ 
    \cline{2-7}
    & & Kitsune \cite{mirsky2018kitsune} & 2018 & AutoEncoder & Online &Packet\\
    &  & MT-FlowFormer \cite{zhao2022mt} & 2022 & Transformer & - & Flow\\
    & & I$^2$RNN \cite{song2022i} & 2022 & I$^2$RNN & - & Packet\\
    &  & ERNN \cite{zhao2022ernn} & 2022 & ERNN & - & Flow\\
    &  & Euler \cite{king2023euler} & 2023 & GNN, RNN & - & Flow \\
    & \multirow{2}{*}{Graph and} & pVoxel \cite{fu2023point} & 2023 & - & - & Packet, Flow\\
    & \multirow{2}{*}{Sequence} & NetVigil \cite{hsieh2024netvigil} & 2024 & E-GraphSage & - & Flow\\
    & \multirow{2}{*}{Neural} & Exosphere \cite{fu2024detecting} & 2024 & CNN & - & Packet\\
    & \multirow{2}{*}{Network} & DFNet \cite{zhao2024effective} & 2024 & DFNet & - & Packet \\
    & & RFH-HELAD \cite{zhong2024rfg} & 2024 & RPGAN, Deep kNN & - & Packet \\
    & & ReTrial \cite{zhao2024retrial} & 2024 &Bayesian Inference & Online & Flow \\
    & &  HEN \cite{wei2024hen} & 2024 & AE-LSTM & - & Packet, Flow\\
    & & TCG-IDS \cite{wu2025tcg} & 2025 & TGN & Online & Flow \\
    & & A-NIDS\cite{zha2025nids} & 2025 & Stacked CTGAN & Online & Flow \\
    & & GTAE-IDS\cite{ghadermazi2025gtae} & 2025 & Graph Transformer & Online & Packet, Flow \\
    \hline

    \multirow{2}{*}{Hybrid} & \multirow{2}{*}{Hybrid} & OWAD \cite{han2023anomaly} & 2024 & Autoencoder & Online & Hybrid \\
    & & FG-CIBGC \cite{ni2025fg} & 2025 & DisenGCN, ICL & - & Hybrid \\
  \bottomrule
\end{tabular}
% \vspace{-5pt}
\end{table}

\subsubsection{Audit Log-based Detectors}
\label{sec:audit}

Audit logs are collected from hosts and thus detectors based on them are basically referred to as HIDS. Audit logs provide fine-grained information through provenance graphs to depict system behaviors. Depending on the learning techniques, audit log-based detectors can be further classified as traditional learning and graph neural network.

\paragraph{Traditional Learning}

Traditional learning-based detectors refer to those that utilize naive machine learning techniques.
StreamSpot \cite{r111} is a clustering-based anomaly detection that tackles challenges in heterogeneity and streaming nature.
Unicorn \cite{r25} is a real-time intrusion detector that efficiently constructs a streaming histogram to represent the history of system executions. The counting results within the histogram are updated immediately if new edges (or events) occur.
DistDet \cite{dong2023distdet} is a distributed detection system that builds host models in the client side, filters false alarms based on their semantics, and derives global models to complement the host models.
Velox \cite{bilot2025sometimes} derives from Orthrus and replaces the complex TGN-based encoder with a simple fully-connected network (FCN), leading to a lightweight and efficient neural network.

\paragraph{Graph Neural Network}
GNN is demonstrated to do well in processing provenance graphs \cite{r24, r31, r28, r29, r30}. 
ProGrapher \cite{r24} extracts temporal-ordered provenance graph snapshots from the ingested logs, and applies whole graph embedding and sequence-based learning to capture rich structural properties of them. The key GNN technique leveraged by ProGrapher is graph2vec.
ShadeWatcher \cite{r29} is a recommendation-guided intrusion detector using provenance graphs. It borrows the recommendation concepts of user-item interactions into security concepts of system entity interactions and analyzes cyber threats in an automated and adaptive manner.
threaTrace \cite{r30} emerges as an online approach dedicated to detecting host-based threats at the node level. Its GNN model is a tailored GraphSAGE \cite{hamilton2017inductive} for learning rich contextual information in provenance graphs.
MAGIC \cite{r31} leverages Graph Attention Network (GAT) \cite{velickovic2017graph} as its graph representation module. MAGIC employs masked graph representation learning to incorporate the capability of pre-training. It can adapt to concept drift with minimal computational overhead, making it applicable to real-world online APT detection.
Flash \cite{r28} is a comprehensive and scalable approach on data provenance graphs to overcome the limitations in accuracy, practicality, and scalability. Flash incorporates a novel adaptation of a GNN-based contextual encoder to encode both local and global graph structures into node embeddings efficiently.
R-caid \cite{goyal2024r} first incorporates root cause analysis into PIDS. Before training GNNs, R-caid links nodes to their root causes to build a new graph, intending to prevent it from mimicry and evasion attacks.
Argus \cite{xu2024understanding} finds the performance of the prior IDS is questionable on large scale. It thus devises a form of discrete temporal graph and uses encoder-decoder unsupervised learning to detect different types of attacks.
TAPAS \cite{zhang2025tapas} leverages a stacked LSTM-GRU model and a task-guided segmentation algorithm to reduce the spatiotemporal dimensions of APT detection, achieving efficient, low-cost, and accurate detection.
In addition to the aforementioned detectors, recent researchers have developed numerous useful tools for better understanding audit logs, such as data visualization analysis tool \cite{liu2025we} and counterfactual-driven attack explanation generator \cite{wu2025provx}.

\subsubsection{Application Log-based Detectors}
\label{sec:application}

Application logs are generated from the installed binaries. Generally, application logs are in the form of natural language text, namely sequence data. It is thus common to introduce sequence-based DL techniques into application log-based DL-IDS.

\paragraph{Traditional Learning}
For traditional learning, Wei et al. \cite{xu2009detecting} propose a general methodology to mine rich semantic information in console logs to detect large-scale system problems. 
Bodik et al. \cite{bodik2010fingerprinting} leverage a logistic regression model on a new and efficient representation of a datacenter's state called fingerprint to detect previously seen performance crises in that datacenter. 
AMOD \cite{dong2018adaptive} uses the SVM HYBRID strategy to filter query annotations from web request logs and then update the stacked generalization detection model to efficiently detect web code injection attacks and obtain malicious queries to update the web application firewall (WAF) library.

\paragraph{Sequence Neural Network}
Due to the similarity between application logs and natural language texts, sequence neural networks such as Recurrent Neural Network \cite{hochreiter1997long} and Transformer \cite{vaswani2017attention,devlin2019bert} are widely employed.
DeepLog \cite{r34} employs LSTM to model system logs as natural language sequences. It is able to automatically learn benign log patterns and detect anomalies when there is a deviation between log patterns and the trained model.
LogRobust \cite{zhang2019robust} finds previous methods do not work well under the close-world assumption and utilizes an attention-based LSTM model to handle unstable log events and sequences.
LogAnomaly \cite{r39} identifies previous studies tend to cause false alarms by using indexes rather than semantics of log templates. Empowered by a novel, simple yet effective method termed template2vec, LogAnomaly is proven to successfully detect both sequential and quantitive log anomalies simultaneously.
LogC \cite{r38} is a new log-based anomaly detection approach with component-aware analysis. It feeds both log template sequences and component sequences to train a combined LSTM model for detecting anomalous logs.
NeuralLog \cite{r44} targets the performance caused by log parsing errors such as out-of-vocabulary words and semantic misunderstandings and employ BERT to perform neural representation.
PLELog \cite{yang2021semi} is a semi-supervised anomaly detection approach that can get rid of time-consuming manual labeling and incorporate the knowledge on historical anomalies.
SpikeLog \cite{qi2023spikelog} adopts a weakly supervised approach to train an anomaly score model, with the objective of handling a more reasonable premise scenario where a large number of logs are unlabeled.
LogCraft \cite{zhang2024end} is an end-to-end unsupervised log anomaly detection framework based on automated machine learning, which mitigates the cost of understanding datasets and makes multiple attempts for building algorithms.
Tweezers \cite{cui2024tweezers} uses a large language model to identify entities and build a relationship graph, and generates embeddings through graph attention network optimization to achieve security incident detection.
LogSer \cite{chai2024log} parses logs by preprocessing parameters, splitting logs, tree parsing, and template merging. It then inputs relevant embeddings into BERT training to detect anomalies, generate reports, and perform incremental updates.
LogDLR \cite{zhou2025logdlr} uses SBERT embeddings and a Transformer autoencoder with domain adversarial training to learn domain-invariant features, detecting anomalies via reconstruction error.

\subsubsection{Network Traffic-based Detectors}
\label{sec:network}
Network traffic comes from communications between hosts across a computer network. It is ruled by network protocols such as Transmission Control Protocol (TCP) and the User Datagram Protocol (UDP) and can be utilized for intrusion detection. Basically, network traffic-based detectors are termed NIDS.

\paragraph{Traditional Learning}
Given the fact that network traffic is usually encrypted for secure communications, feature engineering-guided machine learning is widely applied in NIDS. 
NetPro \cite{li2017netpro} employs traceability reasoning with Merkle Hash Trees and digital signatures to detect direct and indirect MANET routing attacks while preserving node privacy, and outputs a traceability graph to identify malicious nodes and behaviors.
CATH \cite{guo2019cath} is a catastrophe-theory-based approach for DoS detection in software-defined networks (SDNs), which leverages the selection, normalization, and fusion of statistical flow attributes to model network states.
Whisper \cite{fu2021realtime} pays attention to both high accuracy and high throughput by utilizing frequency domain features.
SigML++ \cite{r52} is an extension of SigML for supervised anomaly detection approach. SigML++ employs Fully Homomorphic Encryption and Artificial Neural Network (ANN) for detection, resulting in execution without decrypting the logs.
OADSD \cite{zhang2023real} achieves task independently and has the ability of adapting to the environment over SD-WAN by using On-demand Evolving Isolation Forest.
LtRFT \cite{tang2023ltrft} innovatively introduces Learning-To-Rank scheme for mitigating the low-rate DDoS attacks targeted at flow tables.
AGC \cite{wu2025intrusion} maps the original data into the embedding space through embedding learning to obtain more representative anchor points, thus achieving fine-grained classification of low-quality label data.

\paragraph{Graph and Sequence Neural Network}
In network traffic, packets consist of various contents and their flows can be represented as graphs. As a result, both graph neural network and sequence neural network are adopted in NIDS.
Kitsune \cite{mirsky2018kitsune} is a plug and play NIDS that is allowed to detect attacks efficiently on the local network without supervision. It alleviates the problem that network gateways and router devices simply do not have the memory or processing power. 
MT-FlowFormer \cite{zhao2022mt} is a semi-supervised framework to mitigate the lack of a mechanism for modeling correlations between flows and the requirement of a large volume of manually labeled data.
I$^2$RNN \cite{song2022i} is an incremental and interpretable RNN for encrypted traffic classification, which can be efficiently adapted for incremental traffic types.
ERNN \cite{zhao2022ernn} represents error-resilient RNN, which is a robust and end-to-end RNN model specially designed against network-induced phenomena.
Euler \cite{king2023euler} accelerates the most memory-intensive part, message-passing stage within GNN, with several concurrently-executed replicated GNNs. 
pVoxel \cite{fu2023point} is an unsupervised method that proposes to leverage point cloud analysis to reduce false positives for the previous NIDS such as Whisper and Kitsune without requiring any prior knowledge on the alarms.
NetVigil \cite{hsieh2024netvigil} is specially designed for east-west traffic within data center networks. It utilizes E-GraphSage and contrastive learning techniques to strengthen its resilience.
Exosphere \cite{fu2024detecting} detects flooding attacks by analyzing packet length patterns, without investigating any information in encrypted packets.
DFNet \cite{zhao2024effective} is a DDoS prevention paradigm denoted by preference-driven and in-network enforced shaping.
RFH-HELAD \cite{zhong2024rfg} consists of a $K$ classification model based on a deep neural network and a $K+1$ classification combining GAN and Deep kNN for detecting anomalies in network traffic.
ReTrial \cite{zhao2024retrial} employs an improved graph attention network with Bayesian and EM algorithms to iteratively correct misleading links, enabling robust detection of encrypted malicious traffic.
HEN \cite{wei2024hen} uses SMOTE to enhance data, trains LightGBM, generates explanations via SHAP, trains AE-LSTM to reconstruct SHAP values, sets a threshold from training errors, and marks test traffic with excess errors as attacks for intrusion detection.
TCG-IDS \cite{wu2025tcg} is the first self-supervised temporal contrastive GNN for network intrusion detection, capturing spatiotemporal traffic dependencies with high accuracy and low false alarms.
A-NIDS \cite{zha2025nids} uses a shallow fully connected network for real-time detection and a Stacked CTGAN generator to address catastrophic forgetting and old data storage costs.
GTAE-IDS \cite{ghadermazi2025gtae} uses a graph autoencoder with a Transformer encoder and DNN decoder to learn benign traffic, enabling label-free, near-real-time intrusion detection and new attack identification.

\subsubsection{Hybrid Log-based Detectors}
Based on the above discussions, a natural idea is to combine various types of log data for improving detection capability.
OWAD \cite{han2023anomaly} is a general framework to detect, explain, and adapt to normality shifts in practice. OWAD is validated to be effective in various detection granularity, covering provenance graphs, application logs, and network packets.
FG-CIBGC \cite{ni2025fg} mines syncretic semantics in multi-source logs including audit logs, application logs, and network traffic using LLM under in-context learning, which generates behavior graphs for comprehensive analysis.

\subsection{Attack Investigation}
\label{sec:investigation}

Except for identifying individual intrusive nodes, IDS are supposed to detect the full story of intrusions (a.k.a., attack scenario graphs). This process is referred to as attack investigation, which can be done by directly detecting attack scenario graphs \cite{r26}, or analyzing the causal relations between compromised nodes progressively to construct attack scenario graphs \cite{r69, r60, r68, jiang2025orthrus}. The attack scenario graphs are defined with scenario graphs as follows:
\begin{definition}
\label{def:sg}
    \textit{(Scenario Graph)}. Scenario graph is a subgraph of its given provenance graph, which is constructed by the nodes and edges causally dependent on \textit{nodes of interest}.
\end{definition}
\begin{definition}
\label{def:as}
    \textit{(Attack Scenario Graph)}. Attack scenario graph is a scenario graph where its nodes of interest are compromised nodes.
\end{definition}

\noindent In the past, attack investigation is conducted by forward analysis and backward analysis \cite{hossain2017sleuth}. Forward analysis discovers the influence that nodes of interest will cause and backward analysis traces back how nodes of interest are generated. Benefiting from DL techniques, both forward and backward analysis can be achieved by learning patterns of attack scenario graphs. Furthermore, visual analytics techniques have been widely used to assist security analysts in understanding the causal chain of intrusions \cite{zhang2017survey, zhao2013idsradar}. Table~\ref{tab:investigation} summarizes the overview of attack investigation approaches. Similar to Section~\ref{sec:detection}, we exclude papers \cite{fang2022back, gao2018aiql, ma2017mpi, r56, r76, r65, yang2020uiscope, kwon2018mci, alhanahnah2022autompi, ji2017rain, zhu2023aptshield, hossain2017sleuth, wang2023tesec} slightly relevant to DL for conciseness.

\begin{table}[t]
\centering
  \caption{Overview of attack investigation approaches.}
  \label{tab:investigation}
  \footnotesize
  \begin{tabular}{ccccccccc}
    \toprule
    \textbf{Taxonomy}
    & \textbf{Approach}
    & \textbf{\makecell{Release\\Time}}
    & \textbf{\makecell{Audit\\Log}}
    & \textbf{\makecell{Applica-\\tion Log}}
    & \textbf{\makecell{Base\\Model}}
    & \textbf{\makecell{Starting\\Node}}
    & \textbf{\makecell{Investigation\\Granularity}}
    \\
    \midrule
    Traditional & ProvDetector \cite{r26} & 2020 & \cmark & \xmark & doc2vec & \xmark & Path\\
    Learning & BehaviorBaseline \cite{zhucase} & 2025 & \cmark & \xmark & FastText & \xmark & Path\\
    \hline
    \multirow{2}{*}{Sequence} & ATLAS \cite{r69} & 2021 & \cmark & \cmark & LSTM & \cmark & Graph\\ 
    \multirow{2}{*}{Neural} & LogTracer \cite{r64} & 2022 & \cmark & \cmark & DeepLog & \cmark & Path\\ 
    \multirow{2}{*}{Network} & ConLBS \cite{r62}& 2023 & \cmark & \xmark & Transformer & \cmark & Graph\\
    & AirTag \cite{r60} & 2023 & \cmark & \cmark & BERT & \xmark& Graph\\ 
    \hline
    & Liu et al. \cite{r59} & 2022 & \cmark & \xmark& struc2vec & \cmark & Graph\\ 
    \multirow{2}{*}{Graph} & Karios \cite{r67} & 2023 & \cmark & \cmark & GNN & \xmark & Graph\\
    \multirow{2}{*}{Neural}& TREC \cite{lv2024trec} & 2024 & \cmark & \xmark & GNN & \xmark & Graph \\
    \multirow{2}{*}{Network}& Orthrus \cite{jiang2025orthrus} & 2025 & \cmark & \xmark & UniMP & \xmark & Path\\
    & Slot \cite{qiao2024slot} & 2025 & \cmark & \xmark & GNN & \xmark & Graph \\
    & FeCoGraph \cite{mao2025fecograph} & 2025 & \xmark & \xmark & GCN & \xmark & Graph \\
  \bottomrule
\end{tabular}
\end{table}

\paragraph{Traditional Learning}
Unlike detecting intrusive nodes, attack scenario graphs are complicated and thus are hard to handle by traditional learning methods. ProvDetector \cite{r26} utilizes doc2vec to learn the embedding representation of paths in the provenance graph. Then a density-based detection is deployed to detect abnormal causal paths in the provenance graph. 
BehaviorBaseline \cite{zhucase} presents a novel learning-based anomaly detection method for large-scale provenance graphs. It incorporates dynamic graph processing with adaptive encoding and a tag-propagation framework for real-time detction.

\paragraph{Sequence Neural Network}
Log data is in the form of natural language text or is allowed to be transformed into sequences of events, which facilitates the introduction of sequence neural networks.
ATLAS \cite{r69} is a framework to construct end-to-end attack stories from readily available audit logs, which employs a novel combination of causal analysis and natural language processing. ATLAS exploits LSTM to automatically learn the pattern difference between attack and non-attack sequences.
LogTracer \cite{r64} is an efficient anomaly tracing framework that combines data provenance and system log detection together. An outlier function with an abnormal decay rate is introduced to improve the accuracy.
ConLBS \cite{r62} combines a contrastive learning framework and multilayer Transformer network for behavior sequence classification. 
AirTag \cite{r60} employs unsupervised learning to train BERT directly from log texts rather than relying on provenance graphs. AirTag constructs attack scenario graphs by integrating the detected victim nodes.

\paragraph{Graph Neural Network}

To capture causal relations within graphs, GNN is commonly adopted.
Liu et al. \cite{r59} propose an automated attack detection and investigation method via learning the context semantics of the provenance graph. The provenance graph analyzed by struc2vec captures temporal and causal dependencies of system events.
Kairos \cite{r67} is a practical intrusion detection and investigation tool based on whole-system provenance. Kairos utilizes GNN to analyze system execution history, so that detects and reconstructs complex APTs. It employs a GNN-based encoder-decoder architecture to learn the temporal evolution of provenance graph structure changes and quantify the abnormal degree of each system event.
TREC \cite{lv2024trec} abstracts APT attack investigation problem as a tactics / techniques recognition problem. TREC trains its model in a few-shot learning manner by adopting a Siamese neural network.
Orthurus \cite{jiang2025orthrus} identifies Quality of Attribution as the key factor contributing to whether or not the industry adopts IDS. It first detects malicious hosts using a GNN encoder and then reconstructs the attack path through dependency analysis.
Slot \cite{qiao2024slot}, based on provenance graphs and graph reinforcement learning, uncovers hidden relationships among system behaviors, dynamically adapts to new activities and attack strategies, resists adversarial attacks, and automatically constructs attack chains.
FeCoGraph \cite{mao2025fecograph} directly processes traffic embedding through line graphs to adapt to various GNNs, covering more attack scenarios while protecting data privacy.

\section{Benchmark Datasets}
\label{sec:benchmark}

DL-IDS relies on high-quality data to train an effective model. This section introduces the dimensions of datasets (Section~\ref{sec:dimension}) and some public datasets widely used in DL-IDS (Section~\ref{sec:public}).

\subsection{Dimensions of Datasets}
\label{sec:dimension}

To illustrate the quality of DL-IDS datasets, it is general to use the following dimensions:

\begin{itemize}
\item \textbf{Benign Scenarios}: Benign data should cover benign behaviors and system activities to the greatest extent, enabling DL-IDS to learn patterns of benign behaviors to differentiate malicious behaviors. 
\item \textbf{Malicious Scenarios}: Malicious data ought to incorporate typical attack scenarios while taking into account the diversity of attacks, including short-term and long-term attacks, as well as simple attacks and multi-stage attacks.
\item \textbf{Ground-truth Labels}: Data should be labeled as benign or malicious. For multi-stage attacks, it is useful to indicate the attack type or the attack stage it belongs to. 
\item \textbf{Data Granularities}: Datasets can be in the form of different granularities. The most accepted one is to provide raw log data. Due to copyright concerns, some replicates \cite{r31, r60} merely provide post-processed log data without their processing source codes.
\item \textbf{Operating Systems}: The operating system determines the generalizability of the dataset. The more operating systems a dataset covers and the more common they are, the more comprehensively it can evaluate PIDS performance.
\end{itemize}

\subsection{Public Datasets}
\label{sec:public}

Publicly available datasets bring a lot of convenience to research on DL-IDS. However, some researchers use self-made datasets that are not publicly available, making it difficult for other researchers to reuse their datasets \cite{r128}. To address this issue, we collect and organize some open-source datasets for further studies, which are listed in Table~\ref{tab:datasets}. 

\begin{table}[t]
\caption{Overview of public datasets. W, L, F, A, M, and S represent the operating system of Windows, Linux, FreeBSD, Android, Mac, and supercomputer, respectively.}
\centering
\footnotesize
\label{tab:datasets}
\begin{tabular}{ccccccc}
\toprule
\textbf{Dataset}&\textbf{Release}&\textbf{Size}&\textbf{Scenarios}&\textbf{Label}&\textbf{Format}&\textbf{System}\\
\midrule
     LANL Dataset \cite{kent-2015-cyberdata1} & 2015 & 12 GB  & - & Yes & .txt & W \\
     StreamSpot \cite{r111} & 2016 & 2 GB & 1 & Yes & .tsv & L\\
     AWSCTD \cite{vceponis2018towards} & 2018 & 39 GB & - & No & SQLite & W \\
     DARPA TC E3 \cite{darpatc}& 2018 & 366 GB \cite{griffith2020scalable} & 6 & No & CDM & W, L, F, A\\
     DARPA TC E5 \cite{darpatc} & 2019 & 2,699 GB \cite{griffith2020scalable} & 8 & No & CDM & W, L, F, A\\
     DARPA OpTC \cite{optc} & 2020 & 1,100 GB \cite{r112} & - & No & eCAR & W\\
     Unicorn SC \cite{r25} & 2020 & 147 GB & 2 & Yes & CDM & L\\
     CERT Dataset \cite{glasser2013bridging, Lindauer2020} & 2020 & 87 GB & - & Yes & .csv & W \\
     LogChunks \cite{r106} & 2020 & 24.1 MB & - & Yes & .txt & -\\
     Loghub \cite{r104} & 2020 & 77 GB & - & - & .txt & W, L, M, S\\
     ATLAS \cite{r69} & 2021 & 0.5 GB & 10 & Yes & .txt & W\\
     ATLASv2 \cite{riddle2023atlasv2} & 2023 & 12 & 10 & Yes & .txt & W\\
     ProvSec \cite{r105} & 2023 & - & 11 & Yes & .json & L\\
     AutoLabel \cite{peng2025autolabel} & 2025 & 136 GB & 29 & Yes & .json & L \\ 
\bottomrule
\end{tabular}
\end{table}

LANL Dataset \cite{kent-2015-cyberdata1} is collected within the internal computer network of Los Alamos National Laboratory's corporate. The dataset consists of 58 consecutive days of de-identified data, covering about 165 million events from 12 thousand users. To obtain, its data sources include Windows-based authentication events, process start and stop events, DNS lookups, network flows, and a set of well-defined red teaming events.

StreamSpot dataset \cite{r111} is made up of 1 attack and 5 benign scenarios. The attack scenario exploits a Flash vulnerability and gains root access to the visiting host by visiting a malicious drive-by download URL. The benign scenarios are relevant to normal browsing activity, specifically watching YouTube, browsing news pages, checking Gmail, downloading files, and playing a video game. All the scenarios are simulated through 100 automated tasks with the Selenium RC \cite{selenium}. 

DARPA TC datasets \cite{darpatc} are sourced from the DARPA Transparent Computing (TC) program, identified by the number of engagements from E1 to E5. Among them, DARPA TC E3 is the most widely used. The TC program aims to make current computing systems transparent by providing high-fidelity visibility during system operations across all layers of software abstraction. Unfortunately, DARPA TC datasets are released without labels, and DARPA makes no warranties as to the correctness, accuracy, or usefulness of the datasets.

DARPA Operationally Transparent Cyber (OpTC) \cite{optc} is a technology transition pilot study funded under Boston Fusion Corporate. The OpTC system architecture is based on the one used in TC program evaluation. In OpTC, every Windows 10 endpoint is equipped with an endpoint sensor that monitors post events, packs them into JSON records, and sends them to Kafka. A translation server aggregates the data into eCAR format and pushes them back to Kafka. OpTC scales TC components from 2 to 1,000 hosts. The dataset consists of approximately 1 TB of compressed JSON data in a highly instrumented environment over two weeks.

Unicorn SC \cite{r25} is a dataset specifically designed for APT detection, proposed by Han et al., authors of the Unicorn model. The dataset includes two supply chain scenarios, wget and shell shock, where each scenario lasts for 3 days to simulate the long-term feature of APT attacks, resulting in provenance data containing 125 benign behaviors and 25 malicious behaviors. The data is saved in the form of provenance graphs, describing the causal relationships during the system execution process. 

CERT Dataset \cite{Lindauer2020} is a collection of synthetic insider threat test datasets that provide both background and malicious actor synthetic data. It is developed by the CERT Division, in collaboration with ExactData, LLC, and under sponsorship from DARPA I2O. CERT dataset learned important lessons about the benefits and limitations of synthetic data in the cybersecurity domain and carefully discussed models of realism for synthetic data.

LogChunks \cite{r106} is an application log dataset for build log analysis, containing 797 annotated Travis CI build logs from 80 GitHub repositories and 29 programming languages. These logs are from mature and popular projects, collected through repository, build, and log sampling. Each log in the dataset has manually labeled text blocks of build failure reasons, search keywords, and structural categories, and cross-validated with the original developers with an accuracy of 94.4\%.

Loghub dataset \cite{r104} is a large collection of system log datasets, providing 19 real-world log data from various software systems, including distributed systems, supercomputers, operating systems, mobile systems, server applications, and standalone software. The objective of Loghub is to fill the significant gap between intelligent automated log analysis techniques and successful deployments in the industry. For the usage scenarios of Loghub, about 35\% are anomaly detection, 13\% are log analysis, and 8\% are security. 

ATLAS dataset \cite{r69} implements 10 attacks based on their detailed reports on real-world APT campaigns and generates audit logs in a controlled testbed environment. Among the ten attacks, four are from single host and the rest six are from multiple hosts. All attacks were developed and executed on Windows 7 32-bit virtual machines and took an hour to complete, along with a 24-hour-window audit logs for benign system behaviors.

ATLASv2 dataset \cite{riddle2023atlasv2} enriches the ATLAS dataset with higher quality background noise and additional logging vantage points. In this dataset, two researchers use the victim machines as their primary work stations throughout the course of engagement, instead of depending on automated scripts to generate activity. System logging, in contrast, cover a five-day period, where the first four days simulate normal work days and the fifth day begins with benign activity then trasitions into execution of the corresponding attack.

ProvSec dataset \cite{r105} is created for system provenance forensic analysis. To fulfill data provenance requirements, ProvSec includes the full details of system calls including system parameters. In ProvSec, 11 realistic attack scenarios with real software vulnerabilities and exploits are used and an algorithm to improve the data quality in the system provenance forensics analysis is presented.

AutoLabel dataset \cite{peng2025autolabel} automates fine-grained log labeling by reducing the labeling problem to obtaining an accurate attack subgraph in a provenance graph. Its experiments consist of 29 scenarios, including 25 real CVE vulnerabilities across 12 widely-used applications (spanning 5 programming languages) plus a Sandworm threat simulation by MITRE CTID.

\section{Challenges and Future Directions}
\label{sec:discussion}
After the detailed introduction to the data management stage and the intrusion detection stage, as well as the widely-used benchmark datasets, this section further discusses challenges encountered in existing DL-IDS and summarizes the corresponding visions. These include fundamental resources (Section~\ref{sec:fundamental}), pre-trained large models (Section~\ref{sec:pre-train}), and comprehensive applications (Section~\ref{sec:comprehensive}).

\subsection{Fundamental Resources}
\label{sec:fundamental}

Effective DL-IDS heavily depends on core fundamental resources such as datasets and computing facilities to develop \cite{r110}. Here, we will discuss their challenges one after the other.

\subsubsection{Poor Data Quality}
Existing datasets for DL-IDS may contain errors, inaccuracies, or missing values. This leads to unreliable descriptions of system behaviors that may mislead DL-IDS. For example, in some cases of the DARPA TC dataset, the PROCESS object and its source fail to properly resolve conflicts, resulting in possible incorrect transformation. Besides, the acuity\_level value of the FLOW object is 0, while the value range for this field in other objects is from 1 to 5. Another example could be the LogChunks \cite{r106} dataset. In this dataset, the content describing the failure reasons is possibly incomplete. This is because a chunk in LogChunks only contains a continuous substring of the log text and a failure reason may be described across multiple sections of the log. Moreover, LogChunks neglects the classification of failure reasons like test, compilation, and code inspection errors, which hinders further research from analyzing failure reasons.

Meanwhile, high-quality ground-truth labels are hard to acquire, which is impeded by the contradiction between fine-grained manual labeling and automated label generation. On one hand, for unknown intrusions such as zero-day attacks, it is very labor-intensive for security analysts to correspond each attack scenario to certain log entries, although coarse-grained attack scenarios may have been acquired. The DAPRA TC dataset \cite{darpatc} is a typical example for this. It only provides a ground truth report for attack scenarios, which does not correspond to any specific log entries. Although a few researchers \cite{r30} provide the third-party ground-truth labels that are manually identified by themselves, we empirically find some ambiguities between their ground-truth labels and the official attack scenario report. These ambiguities have an obviously negative effect on DL-IDS, and to some extent, they may even cause the accumulation of errors. On the other hand, the development of automated labeling tools is in an awkward position. The log data is generated based on its given prior knowledge of intrusions \cite{cheng2025tagapt}, whereas the challenge of DL-IDS is to detect zero-day intrusions. This tends the development of such automated tools to be somewhat pointless.

In addition, there are no unified and effective evaluation metrics for DL-IDS \cite{r67}, which further weakens the potential of datasets. For example, precision, recall, F1 score are usually exploited in most studies \cite{r26, r69, r28, r31}, while some papers \cite{r60} propose to use True Positive Rate (TPR) and False Positive Rate (FPR) as evaluation metrics. This makes the comparison experiments usually unfair and hard to tell if the validation is convincing. We also note that in many cases where the percentage of negatives (or malicious log entries) is low, sacrificing FPR can always significantly increase TPR. For example, sacrificing 1,000 false positives for one true positive might only increase FPR by 0.05\%, but would increase TPR by 5\%.

\subsubsection{Insufficient Amount of Data}

Although log data is generated very quickly (e.g., eBay generates 1.2 PB log data per day by 2018 \cite{samuel2018monitoring}), DL-IDS is still facing challenges in insufficient amounts of data. Discounting the above data quality issues such as inaccuracies, the reasons are three-fold:

First, log data has an extremely large number of trivial events, which are proven ineffective and usually removed by graph summarization \cite{r29, r24}. For example, data provenance provides fine-grained information about memory-related events, such as data-to-memory mapping and protection of certain memory addresses. These memory-related events basically do not involve attacks, and unfortunately, are always orthogonal to the existing DL-IDS. However, to ensure the completeness requirement of data provenance and to capture very infrequent but inevitable memory attacks, these memory-related events are still recorded in benchmark datasets. As a result, the usable part of each dataset is rather small for DL-IDS, which can be reflected by the high summarization ratio achieved by graph summarization approaches (e.g., 70\% \cite{r134}). 

The second reason for an insufficient amount of data is the limited dataset representativeness. As observed in Table~\ref{tab:datasets}, most of the datasets have no more than 10 attack scenarios, not to mention that each of these attack scenarios has been carefully chosen by their authors. This limited number of attack scenarios suggests that existing datasets are almost impossible to represent the diversified attack methods, as the number of CVE records has already been over 280,000 \cite{cve}. Furthermore, the existing datasets such as DAPRA TC E3 \cite{darpatc} are collected in a specific experimental environment and may not cover other types of normal system behaviors, and are proven that a significant amount of synthetic data exists \cite{liu2025we}. DARPA TC E5 \cite{darpatc} is unusable for most experiments due to the sparse and error-filled documentation. Unicorn SC \cite{r25} is generated by an idealized simulation of supply chain scenarios, which means many real-world features are prone to be ignored in this dataset. Hence, training DL-IDS on these non-representative datasets could be a disaster for the computer systems that they protect.

Finally, the accessibility of datasets further exacerbates the insufficient data problem. Due to privacy and copyright issues, some datasets may be proprietary or difficult to obtain \cite{r26, wang2023tesec}. Moreover, ProvDetector \cite{r26} conducted a three-month system evaluation in an enterprise environment with 306 hosts and collected benign provenance data of 23 target programs. Yet this dataset has not been made public, rendering it unavailable to improve other DL-IDS and almost all the assessment settings related to ProvDetector are susceptible to inequity.

\subsubsection{Potential Heavy Computation Requirements}

Similar to other DL techniques, DL-IDS also requires a potentially large amount of computing resources to improve their performance. According to \cite{r122}, the generalizability of neural models is proportional to the investment of computing resources. Supposing that the challenge of insufficient data is mitigated and a large volume of log data is available, more computing resources are inevitably required. Besides, we will illustrate in Section~\ref{sec:pre-train} that there are plenty of powerful techniques that have not been introduced in DL-IDS, which will also bring in computation requirements. Unfortunately, acceleration methods like parallel computation and efficient retrieval have not been fully scheduled by the cybersecurity community. An example is that the computation time of Unicorn equipped with one core is proven linear to its workloads \cite{r25}. It is clear that the efficiency of Unicorn, which is not implemented in parallel, will reach the bottleneck as this core does.

\subsubsection{Future Directions}
To conclude, the challenges for DL-IDS in fundamental resources consist of data quality, data volume, and computational overhead. Apart from unintentional errors and non-technical issues in fundamental resources, the research questions that urgently need to be addressed include the contradiction between unaffordable manual labeling and non-generalizable auto-labeling techniques, non-unified benchmark datasets and evaluation metrics, as well as potential heavy computational overheads. Therefore, we summarize the future directions as follows:

\begin{center}
\begin{tcolorbox}[
  colback=gray!10,    % 背景颜色
  colframe=black,        % 边框颜色
  arc=4pt,            % 圆角半径
  boxrule=0.5pt,         % 边框宽度
  left=10pt, right=10pt, top=5pt, bottom=5pt,  % 内边距
  width=0.9\linewidth,
  title=\textbf{Future Directions},
  colbacktitle=black!80
]
\begin{itemize}[leftmargin=10pt]
    \item Developing efficient man-machine interactive log labeling mechanisms and organizing open-source data-sharing platforms accordingly to provide large amounts of high-quality datasets.
    \item Maintaining effective and comprehensive benchmark datasets, accompanied by a unified performance metric framework for a fair comparison.
    \item Investigating parallel or simplified strategies for DL-IDS, and studying their integration with log storage systems to achieve end-to-end acceleration.
\end{itemize}
\end{tcolorbox}
\end{center}

\subsection{Pre-training Theories and Techniques}
\label{sec:pre-train}

In recent years, significant progress has been made by Large Language Models (LLMs) in the field of DL. Their capacity to understand and generate dialogue has been greatly enhanced as the model parameters of LLMs keep rising. T5 \cite{r118}, BERT \cite{devlin2019bert}, GPT \cite{r113}, GPT-4 \cite{r114}, LaMDA \cite{r115}, and LLaMA \cite{r116} are notable examples.

\begin{figure}[t]
    \centering
    \includegraphics[width=\linewidth]{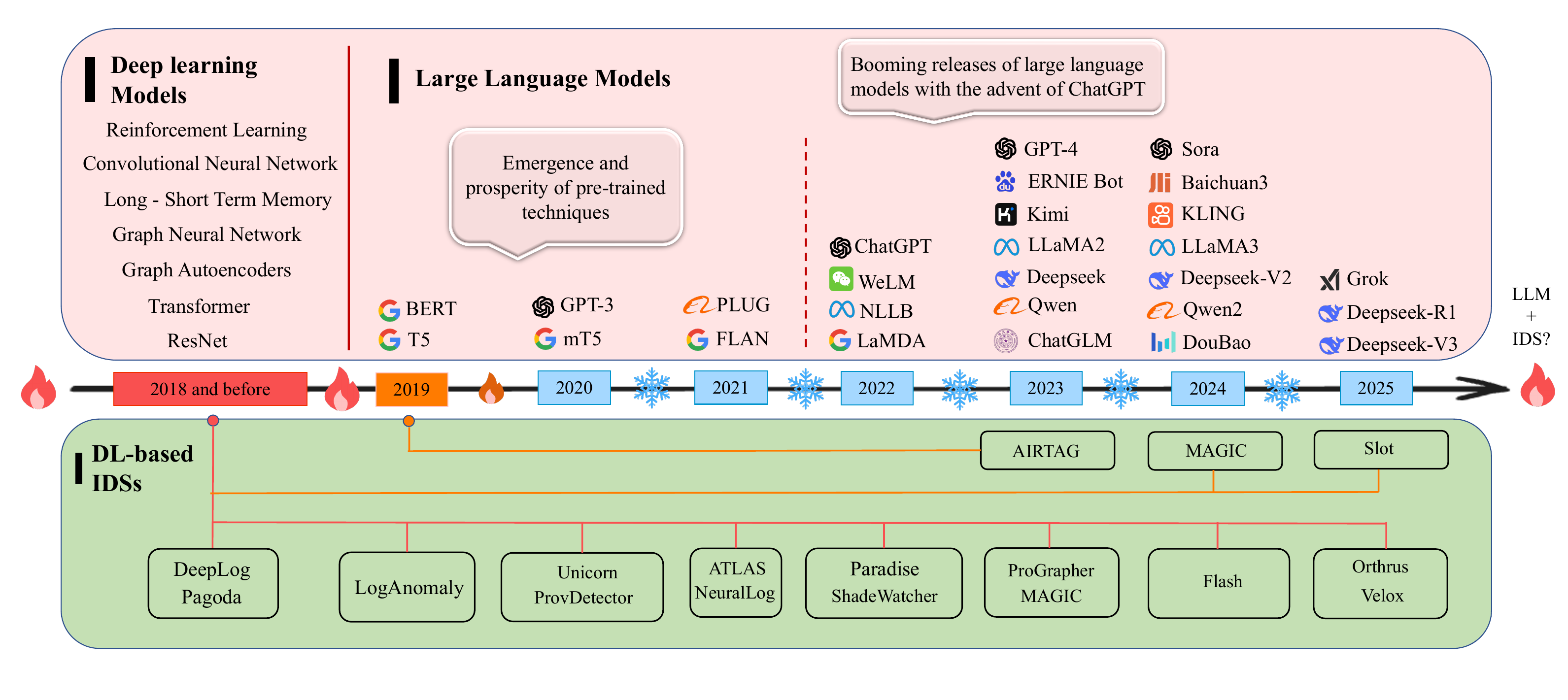}
    \caption{Interactions between DL models and DL-IDS. While DL models proposed before 2019 have already leveraged in DL-IDS, the emerging LLMs (or pre-training theories and thechniques) since 2020 remains underdeveloped in this domain.}
    \label{fig:llm}
\end{figure}

With the development of pre-training techniques, LLMs have been adopted in many fields such as finance \cite{zhao2024revolutionizing}, education \cite{neumann2024llm}, medicine \cite{r123}, and even other domains of cybersecurity \cite{cummins2025llm, hu2024degpt, r124}. In contrast, the adoption of LLMs in DL-IDS is stagnant, as shown in Figure~\ref{fig:llm}. We can observe that LLMs developed at full speed beginning in 2019. Their prosperity, however, has not extended to DL-IDS. Until now, the only two DL-IDS that incorporate pre-training techniques, AirTag \cite{r60} and MAGIC \cite{r31}, still do not make full use of the potential of LLMs. AirTag pre-trains a BERT model on application logs and detects intrusions in terms of embeddings generated by BERT. MAGIC introduces GraphMAE \cite{hou2022graphmae}, a model architecture derived from Graph Autoencoder \cite{kipf2016variational} in 2016 but integrated with the famous masked self-supervised learning method \cite{he2022masked} in 2022, to conduct self-supervised learning on provenance graphs. MAGIC further designs an adapter to apply the pre-trained model in different detection scenarios. Nevertheless, both AirTag and MAGIC can be regarded as preliminary explorations of pre-training techniques. According to the scaling law \cite{kaplan2020scaling}, the performance of LLMs will steadily improve, as the parameters, data, and computation increase. And the reasoning ability of LLMs will suddenly emerge \cite{wei2022emergent}, allowing them to chat with humans smoothly. Such advantageous abilities obviously have not been incorporated into DL-IDS.

Nowadays, some researchers \cite{r119, g2024harnessing, li2024ids, mukherjee2025llm} have started to explore the applications of LLMs on DL-IDS. Yet the theories and techniques of such combination remain challenges. In the following, we will illustrate the identified issues and then point out the future directions.

\subsubsection{Trade-off between Reliability and Generalizability}

The governing concern for the employment of LLMs in DL-IDS is reliability (or explainability). Although offering generalizability, LLMs have long been denounced to have issues with hallucinations \cite{martino2023knowledge, yao2023llm},
privacy \cite{yao2024survey, yang2025privacy, he2025artificial}, overreliance \cite{kim2024m}, and backdoor threats \cite{liu2025backdoor}. These unexplainable and uncontrollable features are an absolute disaster for DL-IDS. For example, when feeding log data to LLMs, they sometimes are prone to hallucinate and provide wrong detection results. Attacks thus successfully bypass the detection facilities and can exfiltrate sensitive data in the victim computer systems. Another example for this is that sensitive information may leak from LLMs. Hui et al. \cite{hui2024pleak} present a prompt leakage attack for LLMs, which is demonstrated to be effective in both offline settings and real-world LLM applications.

\subsubsection{Short of Statistical Log Modeling}

LLMs are developed on the basis of statistical language modeling \cite{rosenfeld2000two, jozefowicz2016exploring}, which is not insufficiently studied for log data. The statistical modeling of natural language can be traced back to the early 1950s when Shannon pioneered the technique of predicting the next element of natural language text \cite{shannon1948mathematical} and discussed the n-gram model for English \cite{shannon1951redundancy}. After that, as machine learning came into view of the NLP research communities, language modeling flourished, and many models such as TreeBank \cite{marcus1993building}, word2vec \cite{mikolov2013efficient, mikolov2013distributed} and LSTM \cite{hochreiter1997long} were proposed. Over decades, researchers in NLP have gained solid knowledge of language modeling, whose interests gradually shifted to efficiency. An epoch-making model, Transformer \cite{vaswani2017attention}, was presented using the multi-head self-attention mechanism to fulfill parallel computing, which was widely exploited in popular pre-trained models such as BERT \cite{devlin2019bert} and GPT \cite{r114} afterward. It is evident that the success of LLMs comes from the prolonged studies on statistical language modeling.

\begin{table}[t]
    \centering
    \footnotesize
    \caption{Comparison of research advances in statistical modeling of various data. ``NL'', ``PL'' and ``FL'' represent Natural Language, Programming Language, and Formal Language, respectively. Note that PL is a type of FL.}
    \begin{tabular}{ccccc}
        \toprule
         Data & Form & Content Generation Rules & Statistical Modeling Studies & Pre-training\\
         \midrule
         Text & NL & Grammar, pragmatics, semantics, etc & \cite{rosenfeld2000two, jozefowicz2016exploring, shannon1951redundancy, marcus1993building} & well-done \\
         Speech & NL & Text rules (see above) and phonetics & \cite{nussbaum2025understanding, kersta1960human} & well-done \\
         Source code & PL & Lexical and syntactic definitions & \cite{hindle2016naturalness, ray2016naturalness, allamanis2018survey} & well-done \\
         \midrule
         Log & NL + FL & Log template defined by developers & future work & underdeveloped\\
        \bottomrule
    \end{tabular}
    \label{tab:data_modeling}
\end{table}

Unfortunately, there are almost no research efforts on statistical modeling of log data, resulting in pre-training techniques of DL-IDS remaining underdeveloped. By contrast, as illustrated in Table~\ref{tab:data_modeling}, the statistical modeling studies of other types of data have already started. Hindle et al. \cite{hindle2016naturalness} demonstrate that the source code is very repetitive and predictable, and, in fact, even more so than natural language. Driven by such statistical modeling conclusion, DL-based source code applications \cite{xu2023xastnn, xu2024dsfm, li2024trident, li2022competition, svyatkovskiy2020intellicode, feng2020codebert, gu2025effectiveness} such as code generation and code clone detection flourish, many of which have already becomes common applications in LLMs. Similar cases can be found for speech data, whose applications are like text to speech \cite{ren2019fastspeech, guo2024voiceflow, park2025nanovoice} and speech recognition \cite{baevski2021unsupervised}.

We argue that log data is also created by humans, similar to text, speech, and source code. It is generated according to developer-defined log templates, with a form of both natural language (e.g., application logs) and formal language (e.g., data provenance in CDM format). Given the fact that natural language (e.g., text and speech) and formal language (e.g., source code) both exhibit positive performance in pre-training, log data urgently demands statistical modeling achievements to facilitate its pre-training research. Although several works \cite{michael2020forensic, r6} have discussed the features of log data, they are orthogonal to the explainable combination of DL and IDS. Compared with the other data types, challenges in statistical log modeling, for instance, may lie in that logs are extremely long and detailed for reliable purposes. It is very common that the length of one single log entry is the same as that of one paragraph in natural language texts. These challenges happen to be the shortcomings of LLMs - not being able to handle long text and not being trustworthy in generated contents.

\subsubsection{Future Directions} According to the scaling laws \cite{kaplan2020scaling} and emergent abilities theory \cite{wei2022emergent}, as the model size continues to grow, the performance of DL-IDS will increase simultaneously. Thus, increasing the amount of model parameters will be an inevitable trend for DL-IDS. The underlying research questions include the strategies for incorporating existing LLMs in intrusion detection, since it is infeasible to directly leverage unreliable LLMs to detect intrusions, and the theories and techniques for modeling long and detailed log data. We summarize the future directions as follows:

\begin{center}
\begin{tcolorbox}[
  colback=gray!10,    % 背景颜色
  colframe=black,        % 边框颜色
  arc=4pt,            % 圆角半径
  boxrule=0.5pt,         % 边框宽度
  left=10pt, right=10pt, top=5pt, bottom=5pt,  % 内边距
  width=0.9\linewidth,
  title=\textbf{Future Directions},
  colbacktitle=black!80
]
\begin{itemize}[leftmargin=10pt]
    \item Investigating how and where to introduce LLMs into DL-IDS like \cite{ni2025fg}, with the objective of balancing the generalizability provided by LLMs and the reliability required by DL-IDS.
    \item Exploring fundamental statistical modeling theories for log data. On this basis, designing pre-training frameworks for log data and its downstream tasks such as steps within the workflow of DL-IDS (see Section~\ref{sec:workflow}).
\end{itemize}
\end{tcolorbox}
\end{center}

\subsection{Comprehensive Applications and Scenarios}
\label{sec:comprehensive}

DL-IDS possess abilities that the traditional IDS lack, or are difficult to realize, such as generalizability for zero-day attacks and modeling ability for complicated downstream tasks. We will elaborate on the possible new-style applications and discuss the challenges in and introduced by them.

\subsubsection{Limited Forward and Backward Tracing Scope}

Forward tracing and backward tracing are employed in attack investigation, as illustrated in Section~\ref{sec:investigation}. Under traditional settings, the forward tracing analyzes the influence a symptom node would have on the victim computer system, and the backward tracing discovers the starting node where the vulnerabilities exist \cite{r10}. 

We argue that the existing tracing scope is too limited to handle increasingly complicated intrusions and DL-IDS can be defined more broadly. In addition to investigating scenario graphs of intrusions, DL-IDS are supposed to further investigate why these intrusions occur and how to hold back them. The broader definition introduces more downstream tasks that would be difficult to accomplish without the assistance of DL techniques. Based on Definition~\ref{def:i}, we reformulate the definition of intrusion in a broad sense for DL-IDS as follows:
\begin{definition}
\label{def:gi}
    \textit{(Generalized Intrusion)}. Generalized intrusion is the malicious attempts against a computer, a network, or the corresponding security facilities, whose attributes encompass not only itself but also its underlying root causes and the relevant control measures.
\end{definition}
In this way, the detection of DL-IDS has been extended to the broadly defined intrusions, including their attributes of both root causes and control measures. When executing backward tracing analysis, DL-IDS are not only required to detect the starting symptom nodes of intrusions, but also required to find the root causes of these symptom nodes (i.e., vulnerabilities in source codes). In the forward tracing analysis, except for detecting the symptom nodes affected by intrusions, DL-IDS should perform an in-depth analysis to discover the potentially compromised nodes and provide control measures for handling intrusions.

Thankfully, several pioneering works have studied similar problems \cite{r127, r120}. In AiVl \cite{r127}, algorithms to bridge log entries and program models are developed using dynamic-static program analysis. Root causes for the exploited vulnerabilities are capable of directly deriving from intrusion detection. Pedro et al. \cite{r120} investigate detection and mitigation methods for DDoS attacks, aiming to control them immediately. Additionally, semi-automated adaptive network defense (SAND) \cite{chen2022sand} leverages SDN to dynamically generate and deploy defense rules. We note that these research attempts are all based on heuristics, either using pre-defined rules to generate root causes, or developing control measures for specific intrusions. Thus, there is a substantial need to introduce advanced DL techniques to this problem.

\subsubsection{Concerns about Data-driven Adversarial Attacks}

To validate the detection performance, DL-IDS commonly idealize the experimental data in their threat model. Such idealization, however, leaves DL-IDS with weaknesses that could be exploited by invaders. For example, a common assumption is that no attacks are considered to compromise the security of the log collection systems \cite{r28, hassan2019nodoze, r31, r25}, namely log data utilized in DL-IDS is absolutely harmless. But as attacks become more stealthy and complicated, it is impossible to satisfy such an assumption apparently. When DL-IDS encounter intentional data poisoning attacks, prediction backdoors could be easily planted as persistent vulnerabilities.

The robustness of DL-IDS is also challenged by data-driven evasion attacks. To evade the detection, the malicious behaviors usually mimic the benign ones (a.k.a., mimicry attacks), making them hard to be detected. By early 2002, David et al. \cite{wagner2002mimicry} have indicated the danger of mimicry attacks on HIDS. Recently, researchers have started to investigate mimicry attacks on DL-IDS \cite{mukherjee2023evading, r130, liu2022vulnergan} and their studies all present effective evasion of detection. From a study \cite{chakraborty2018adversarial}, DL-IDS can be even plagued by a trivial perturbation in log data. Aware of this issue, R-caid \cite{goyal2024r} proposes to embed root causes into the detection model for countering adversarial attacks. However, as noted in recent work \cite{mukherjee2023evading, r130, goyal2024r}, data-driven attacks still remain a major challenge for DL-IDS.

\subsubsection{Underexplored Promising Scenarios}

While DL-IDS show excellent performance in the protection of computer and network systems recently, there are still many promising scenarios for DL-IDS that have not been explored sufficiently. 

Mobile edge computing (MEC) \cite{mao2017survey, abbas2017mobile, li2022defensive} is a typical scenario. In the MEC environment, mobile computing, network control, and storage are pushed at the network edges so as to enable computation-intensive tasks at the resource-limited devices. At the network edges, devices such as Unmanned Aerial Vehicles (UAVs) and New Energy Vehicles (NEVs) usually lack computing power and security facilities, making it difficult to prevent them from intrusions \cite{shrestha2021machine}. In the meantime, containerized deployment has become one of the dominant ways to deploy microservices. Detecting intrusions on containers is thus of great importance, for which ReplicaWatcher \cite{r128} is a representative work with a special design for microservices. Additionally, industrial networks are characterized by high fidelity, stability, and real-time responsiveness \cite{knapp2024industrial}, leading to challenges in adapting DL-IDS to their infrastructures.

\subsubsection{Future Directions}
Although there has been plenty of research on DL-IDS, many applications and scenarios remain underdeveloped. DL-IDS are sought to be more broadly defined and applied. Based on the above discussion, we briefly summarize the future directions as follows:

\begin{center}
\begin{tcolorbox}[
  colback=gray!10,    % 背景颜色
  colframe=black,        % 边框颜色
  arc=4pt,            % 圆角半径
  boxrule=0.5pt,         % 边框宽度
  left=10pt, right=10pt, top=5pt, bottom=5pt,  % 内边距
  width=0.9\linewidth,
  title=\textbf{Future Directions},
  colbacktitle=black!80
]
\begin{itemize}[leftmargin=10pt]
    \item Extending the scope of forward tracing and backward tracing to intrusions in a broad sense, so that generating root causes and control measures for the broadly defined intrusions.
    \item Understanding data-driven adversarial attacks such as data poisoning attacks and mimicry attacks for devising more robust DL-IDS.
    \item Applying DL-IDS widely in more underexplored promising scenarios, and if possible, implementing unified frameworks for them.
\end{itemize}
\end{tcolorbox}
\end{center}

\section{Conclusion}
\label{sec:conclusion}

The DL techniques bring reform to IDS, whose generalizability enables them to detect intrusions that have never been encountered before. Recognizing that the IDS development over the past decade primarily comes from DL-IDS, this survey revisits the common workflow for DL-IDS, elaborates each module in the workflow, and taxonomizes the research papers innovatively based on their DL techniques. Publicly available datasets for stimulating future research are introduced subsequently. In addition, from the perspective of DL, this survey digs deep into the potential challenges, emerging trends, and future directions for DL-IDS. The discussions suggest to us that DL-IDS are, fascinatingly, in an underdeveloped state. We hope that this survey can somewhat inspire current researchers and facilitate future investigations on DL-IDS.

\begin{acks}
This research is sponsored in part by the NSFC program (No. 6212780016 and No. 62021002).
\end{acks}

\bibliographystyle{ACM-Reference-Format}
\bibliography{main}

% \appendix
% \section{Research Methods}

\end{document}